\documentclass[11pt]{article}

\usepackage[dvips]{graphicx}
\usepackage{float}
\usepackage{epsfig}
\usepackage{ulem}
\usepackage{latexsym,amsmath,amsfonts,amssymb}
\usepackage[latin1]{inputenc}
\usepackage{rotating}
\usepackage[american]{babel}
\usepackage[dvips]{graphicx}
\usepackage{bbm}
\usepackage{color}
\usepackage{slashed}
\usepackage{lscape}

\usepackage{hyperref}

\def\im{Invent. Math.}

\def\a{\alpha}
\def\b{\beta}
\def\c{\gamma}
\def\d{\delta}
\def\f{\phi}               
\def\vf{\varphi}  
\def\tvf{\tilde{\varphi}}
\def\vp{\varphi}
\def\g{\gamma}
\def\h{\eta}
\def\j{\psi}
\def\k{\kappa}                    
\def\l{\lambda}
\def\m{\mu}
\def\n{\nu}
\def\o{\omega}  \def\w{\omega}

\def\q{\theta}  \def\th{\theta}                  
\def\r{\rho}                                     
\def\s{\sigma}                                   
\def\t{\tau}
\def\u{\upsilon}
\def\x{\xi}
\def\z{\zeta}
\def\pt{\tilde{\varphi}}

\def\lab{\label}
\def\6{\partial}
\def\wg{\wedge}
\def\bpsi{\bar{\psi}}
\def\bt{\bar{\theta}}
\def\bvf{\bar{\varphi}}

\DeclareMathOperator{\tr}{tr}

\newcommand{\be}{\begin{equation}}
\newcommand{\ee}{\end{equation}}
\newcommand{\beq}{\begin{equation}}
\newcommand{\eeq}{\end{equation}}
\newcommand{\bea}{\begin{eqnarray}}
\newcommand{\eea}{\end{eqnarray}}
\newcommand{\nn}{\nonumber}

\newcommand{\ba}{\begin{eqnarray}}
\newcommand{\ea}{\end{eqnarray}}

\newcommand{\beqs}{\begin{eqnarray}}
\newcommand{\eeqs}{\end{eqnarray}}
\newcommand{\bal}{\begin{aligned}}
\newcommand{\eal}{\end{aligned}}

\begin{document}
\baselineskip=15.5pt
\pagestyle{plain}
\setcounter{page}{1}

\def\del{{\partial}}
\def\vev#1{\left\langle #1 \right\rangle}
\def\cn{{\cal N}}
\def\co{{\cal O}}


\def\IC{{\mathbb C}}
\def\IR{{\mathbb R}}
\def\IZ{{\mathbb Z}}
\def\RP{{\bf RP}}
\def\CP{{\bf CP}}
\def\Poincare{{Poincar\'e }}
\def\tr{{\rm tr}}
\def\tp{{\tilde \Phi}}

\def\TL{\hfil$\displaystyle{##}$}
\def\TR{$\displaystyle{{}##}$\hfil}
\def\TC{\hfil$\displaystyle{##}$\hfil}
\def\TT{\hbox{##}}
\def\HLINE{\noalign{\vskip1\jot}\hline\noalign{\vskip1\jot}}
\def\seqalign#1#2{\vcenter{\openup1\jot
   \halign{\strut #1\cr #2 \cr}}}
\def\lbldef#1#2{\expandafter\gdef\csname #1\endcsname {#2}}
\def\eqn#1#2{\lbldef{#1}{(\ref{#1})}%
\begin{equation} #2 \label{#1} \end{equation}}
\def\eqalign#1{\vcenter{\openup1\jot
     \halign{\strut\span\TL & \span\TR\cr #1 \cr
    }}}

\def\eno#1{(\ref{#1})}
\def\half{\frac{1}{2}}



\def\ads{{\it AdS}}
\def\adsp{{\it AdS}$_{p+2}$}
\def\cft{{\it CFT}}

\newcommand{\ber}{\begin{eqnarray}}
\newcommand{\eer}{\end{eqnarray}}

\newcommand{\beqar}{\begin{eqnarray}}
\newcommand{\cN}{{\cal N}}
\newcommand{\cO}{{\cal O}}
\newcommand{\cA}{{\cal A}}
\newcommand{\cT}{{\cal T}}
\newcommand{\cF}{{\cal F}}
\newcommand{\cC}{{\cal C}}
\newcommand{\cR}{{\cal R}}
\newcommand{\cW}{{\cal W}}
\newcommand{\eeqar}{\end{eqnarray}}
\newcommand{\tht}{\thteta}
\newcommand{\lm}{\lambda}\newcommand{\Lm}{\Lambda}


\newcommand{\nonu}{\nonumber}
\newcommand{\oh}{\displaystyle{\frac{1}{2}}}
\newcommand{\dsl}
   {\kern.06em\hbox{\raise.15ex\hbox{$/$}\kern-.56em\hbox{$\partial$}}}
\newcommand{\id}{i\!\!\not\!\partial}
\newcommand{\as}{\not\!\! A}
\newcommand{\ps}{\not\! p}
\newcommand{\ks}{\not\! k}
\newcommand{\D}{{\cal{D}}}
\newcommand{\dv}{d^2x}
\newcommand{\Z}{{\cal Z}}
\newcommand{\N}{{\cal N}}
\newcommand{\Dsl}{\not\!\! D}
\newcommand{\Bsl}{\not\!\! B}
\newcommand{\Psl}{\not\!\! P}

\newcommand{\eeqarr}{\end{eqnarray}}
\newcommand{\ZZ}{{\rm \kern 0.275em Z \kern -0.92em Z}\;}


\def\del{{\delta^{\hbox{\sevenrm B}}}} \def\ex{{\hbox{\rm e}}}
\def\azb{A_{\bar z}} \def\az{A_z} \def\bzb{B_{\bar z}} \def\bz{B_z}
\def\czb{C_{\bar z}} \def\cz{C_z} \def\dzb{D_{\bar z}} \def\dz{D_z}
\def\im{{\hbox{\rm Im}}} \def\mod{{\hbox{\rm mod}}} \def\tr{{\hbox{\rm Tr}}}
\def\ch{{\hbox{\rm ch}}} \def\imp{{\hbox{\sevenrm Im}}}
\def\trp{{\hbox{\sevenrm Tr}}} \def\vol{{\hbox{\rm Vol}}}
\def\rl{\Lambda_{\hbox{\sevenrm R}}} \def\wl{\Lambda_{\hbox{\sevenrm W}}}
\def\fc{{\cal F}_{k+\cox}} \def\vev{vacuum expectation value}
\def\nodiv{\mid{\hbox{\hskip-7.8pt/}}}
\def\ie{{\em i.e.}}
\def\ie{\hbox{\it i.e.}}

\def\CC{{\mathchoice
{\rm C\mkern-8mu\vrule height1.45ex depth-.05ex
width.05em\mkern9mu\kern-.05em}
{\rm C\mkern-8mu\vrule height1.45ex depth-.05ex
width.05em\mkern9mu\kern-.05em}
{\rm C\mkern-8mu\vrule height1ex depth-.07ex
width.035em\mkern9mu\kern-.035em}
{\rm C\mkern-8mu\vrule height.65ex depth-.1ex
width.025em\mkern8mu\kern-.025em}}}

\def\RR{{\rm I\kern-1.6pt {\rm R}}}
\def\NN{{\rm I\!N}}
\def\ZZ{{\rm Z}\kern-3.8pt {\rm Z} \kern2pt}
\def\IB{\relax{\rm I\kern-.18em B}}
\def\ID{\relax{\rm I\kern-.18em D}}
\def\II{\relax{\rm I\kern-.18em I}}
\def\IP{\relax{\rm I\kern-.18em P}}
\newcommand{\CS}{{\scriptstyle {\rm CS}}}
\newcommand{\CSs}{{\scriptscriptstyle {\rm CS}}}
\newcommand{\rc}{\nonumber\\}
\newcommand{\bear}{\begin{eqnarray}}
\newcommand{\eear}{\end{eqnarray}}

\newcommand{\LL}{{\cal L}}

\def\mani{{\cal M}}
\def\calo{{\cal O}}
\def\calb{{\cal B}}
\def\calw{{\cal W}}
\def\calz{{\cal Z}}
\def\cald{{\cal D}}
\def\calc{{\cal C}}

\def\to{\rightarrow}
\def\ele{{\hbox{\sevenrm L}}}
\def\ere{{\hbox{\sevenrm R}}}
\def\zb{{\bar z}}
\def\wb{{\bar w}}
\def\nodiv{\mid{\hbox{\hskip-7.8pt/}}}
\def\menos{\hbox{\hskip-2.9pt}}
\def\dr{\dot R_}
\def\drr{\dot r_}
\def\ds{\dot s_}
\def\da{\dot A_}
\def\dga{\dot \gamma_}
\def\ga{\gamma_}
\def\dal{\dot\alpha_}
\def\al{\alpha_}
\def\cl{{closed}}
\def\cls{{closing}}
\def\vev{vacuum expectation value}
\def\tr{{\rm Tr}}
\def\to{\rightarrow}
\def\too{\longrightarrow}


\def\a{\alpha}
\def\b{\beta}
\def\c{\gamma}
\def\d{\delta}
\def\e{\epsilon}           
\def\F{\Phi}
\def\f{\phi}               
\def\vf{\varphi}  \def\tvf{\tilde{\varphi}}
\def\vp{\varphi}
\def\g{\gamma}
\def\h{\eta}
\def\j{\psi}
\def\k{\kappa}                    
\def\l{\lambda}
\def\m{\mu}
\def\n{\nu}
\def\o{\omega}  \def\w{\omega}
\def\q{\theta}  \def\th{\theta}                  
\def\r{\rho}                                     
\def\s{\sigma}                                   
\def\t{\tau}
\def\u{\upsilon}
\def\x{\xi}
\def\X{\Xi}
\def\z{\zeta}
\def\pt{\tilde{\varphi}}
\def\lab{\label}
\def\6{\partial}
\def\wg{\wedge}
\def\atanh{{\rm arctanh}}
\def\bpsi{\bar{\psi}}
\def\bt{\bar{\theta}}
\def\bvf{\bar{\varphi}}

%



\newfont{\namefont}{cmr10}
\newfont{\addfont}{cmti7 scaled 1440}
\newfont{\boldmathfont}{cmbx10}
\newfont{\headfontb}{cmbx10 scaled 1728}





\newcommand{\re}{\,\mathbb{R}\mbox{e}\,}
\newcommand{\hyph}[1]{$#1$\nobreakdash-\hspace{0pt}}
\providecommand{\abs}[1]{\lvert#1\rvert}
\newcommand{\Nugual}[1]{$\mathcal{N}= #1 $}
\newcommand{\sub}[2]{#1_\text{#2}}
\newcommand{\partfrac}[2]{\frac{\partial #1}{\partial #2}}
\newcommand{\bsp}[1]{\begin{equation} \begin{split} #1 \end{split} \end{equation}}
\newcommand{\calF}{\mathcal{F}}
\newcommand{\calO}{\mathcal{O}}
\newcommand{\calM}{\mathcal{M}}
\newcommand{\calV}{\mathcal{V}}
\newcommand{\bbZ}{\mathbb{Z}}
\newcommand{\bbC}{\mathbb{C}}
\newcommand{\cK}{{\cal K}}

\newcommand{\Thq}{\Theta\left(\r-\r_q\right)}
\newcommand{\Dq}{\d\left(\r-\r_q\right)}
\newcommand{\kten}{\kappa^2_{\left(10\right)}}
\newcommand{\pbi}[1]{\imath^*\left(#1\right)}
\newcommand{\ho}{\hat{\omega}}
\newcommand{\tth}{\tilde{\th}}
\newcommand{\tf}{\tilde{\f}}
\newcommand{\tj}{\tilde{\j}}
\newcommand{\tw}{\tilde{\omega}}
\newcommand{\tz}{\tilde{z}}
\newcommand{\prj}[2]{(\partial_r{#1})(\partial_{\j}{#2})-(\partial_r{#2})(\partial_{\j}{#1})}
\def\atanh{{\rm arctanh}}
\def\sech{{\rm sech}}
\def\csch{{\rm csch}}
\allowdisplaybreaks[1]

\def\red{\textcolor[rgb]{0.98,0.00,0.00}}

\newcommand{\Dan}[1] {{\textcolor{blue}{#1}}}

\numberwithin{equation}{section}

\newcommand{\Tr}{\mbox{Tr}}    


%

\setcounter{footnote}{0}
\renewcommand{\theequation}{{\rm\thesection.\arabic{equation}}}

\begin{titlepage}

\begin{center}
\Large \bf  Holographic Flows in non-Abelian T-dual Geometries.
\end{center}
\vskip 0.2truein
\begin{center}
Niall T. Macpherson$^{a,}$\footnote{niall.macpherson@mib.infn.it}, 
Carlos N\'u\~nez$^{b,}$\footnote{c.nunez@swansea.ac.uk}, 
Daniel C. Thompson$^{c,}$\footnote{daniel.thompson@vub.ac.be} and
S. Zacar\'ias $^{d,}$\footnote{szacarias@fisica.ugto.mx }

\vspace{.2in}
{\it $a$:  Dipartimento di Fisica, Universit\`a di Milano--Bicocca, I-20126 Milano, Italy\\
    and\\
    INFN, sezione di Milano--Bicocca. }\\
\vspace{.2in}
{\it $b$: Department of Physics, Swansea University\\
 Singleton Park, Swansea SA2 8PP, United Kingdom.}\\
\vspace{0.2in}
{\it $c$: Theoretische Natuurkunde, Vrije Universiteit Brussel,
\& The International Solvay Institutes,
Pleinlaan 2, B-1050 Brussels, Belgium.
}\\
\vspace{0.2in}
{\it $d$:  Department of Nuclear and Particle Physics, Faculty of Physics, University of Athens,\\  Athens 15784, Greece\\
and\\
Departamento de F\'{\i}sica, Divisi\'{o}n de Ciencias e Ingenier\'{\i}as, Campus Le\'{o}n, Universidad de Guanajuato, Loma del Bosque No. 103 Col. Lomas del Campestre, C.P. 37150, Le\'{o}n, Guanajuato, M\'{e}xico.}
\vskip 5mm

\vspace{0.2in}
\end{center}
\vspace{0.2in}
\centerline{\bf Abstract:}

We use non-Abelian T-duality to construct new ${\cal N}=1$ 
solutions of type IIA supergravity (and their M-theory lifts) that interpolate between $AdS_5$ 
geometries.  We initiate a study of the holographic 
interpretation of these backgrounds as 
RG flows between conformal fixed points.   Along the way 
we give an elegant formulation of non-Abelian T-duality
when acting on a wide class of backgrounds, 
including those corresponding to such flows, in terms of their $SU(2)$ structure.

\end{titlepage}
\setcounter{footnote}{0}

\tableofcontents

\setcounter{footnote}{0}
\renewcommand{\theequation}{{\rm\thesection.\arabic{equation}}}

 \section{Introduction} 
 
 The great success of the $AdS/CFT$ conjecture \cite{Maldacena:1997re} in giving gravitational descriptions of super-conformal field theories (SCFTs) naturally begs the question of how it can be extended to Quantum Field Theories (QFTs) that are not fixed points.   In that case one expects that a renormalisation group flow is encoded holographically by modifying the radial behaviour, or warp factors, of the space time geometry.    Early examples of holographic RG flows are found in  \cite{Itzhaki:1998dd,
 Girardello:1998pd,Witten:1998zw,Freedman:1999gp,
Klebanov:2000hb, Maldacena:2000yy}.
  
 More precisely, one could consider a UV fixed point described by a (UV)-CFT 
and trigger a flow by giving a vacuum expectation value (VEV) or 
by deforming with a relevant operator. In the IR, one can 
again encounter a  fixed point defined by an (IR)-CFT.   
If both the IR and UV theories admit holographic duals, 
it is reasonable to expect a holographic description of 
the entire RG flow.  Whilst one might expect the IR CFT to have 
less symmetries, this need not be the case.
Indeed,  
one can encounter situations in which the IR fixed point has an enhanced or accidental  symmetry. 
 Two contrasting examples demonstrate this well and  will be central to our present interests; these are the Klebanov Witten flow  \cite{Klebanov:1998hh} and the Klebanov Murugan flow \cite{Klebanov:2007us}. 
  
 The Klebanov Witten (KW) flow starts with a UV fixed point given by an ${\cal N}=2$ SCFT  with $U(N)\times U(N)$ gauge group and matter in bifundamental hyper multiplets. In ${\cal N}=1$ language this theory has four chiral multiplets $A_i$, $B_i$ with $i=1,2$ coming from the hypers and two adjoint scalars $\Phi$, $\tilde{\Phi}$ coming from the ${\cal N}=2$
gauge multiplets.  The super-potential, 
 \begin{equation}
 {\cal W}_{UV} = g \Tr  \left( \Phi A_i B_i  + \tilde{\Phi} B_i A_i \right) \ , 
 \end{equation}
  can be deformed with a relevant operator ${\cal W}^\prime$, 
giving a mass to the adjoint scalars 
 \be
  {\cal W}^\prime=  m \Tr\left( \Phi^2 - \tilde \Phi^2 \right) \ , 
 \ee
 and, in the IR, once we have integrated out these fields this produces the 
super-potential 
 \be\label{eq:WT11}
 {\cal W}_{IR} =   -\frac{g^2}{2m} \Tr \left( A_1 B_1 A_2 B_2 - B_1 A_1 B_2 A_2 \right) \ ,
 \ee 
 that defines the   ${\cal N}=1$ $U(N)\times U(N)$ gauge theory obtained by placing D3 branes at the tip of the cone over the homogenous space $T^{1,1}$ constructed in \cite{Klebanov:1998hh}.   The gravitational description of the UV fixed point is $AdS_5 \times S^5/\mathbb{Z}_2$ and that of the IR is $AdS_5 \times T^{1,1}$.  Giving a supergravity description of the whole flow has proven to be a challenging enterprise and whilst  an ansatz for a gravitational solution describing 
this flow was proposed in \cite{Halmagyi:2004jy}, it involves PDE's and lacks any known analytic solution.

The Klebanov Murugan (KM) flow \cite{Klebanov:2007us}, 
provides a contrasting behaviour in which symmetries become enhanced in the IR.  In this case, 
one begins in the {\em UV} with the ${\cal N}=1$ theory defined by the super-potential eq.~\eqref{eq:WT11} and triggers a flow by giving a VEV $B_2 = a \mathbb{I}$ (with $A_i=B_1=0$).  This evidentially breaks the global symmetries of the theory down, in fact to $SU(2)\times U(1)^2$, and since $\det B_2 \neq 0$ this can be though of as putting the theory on its Baryonic branch.  Remarkably however one sees that the IR 
super-potential becomes, 
 \be\label{eq:WNeq4}
 {\cal W}_{IR} =  \frac{g^2 a}{2m}  \Tr \left( B_1 [ A_2 , A_1]   \right) \ ,
 \ee
 which is nothing more than that of ${\cal N}=4$ SYM.   Geometrically this procedure corresponds to placing the stack of D3 branes at a specific (non-smeared) point in the finite sized $S^2$ at the tip of the resolved conifold. In this case the geometry describing 
the entire flow can be 
written analytically  and the emergence of    ${\cal N}=4$ SYM is obtained again as   an $AdS_5\times S^5$ throat near the location of the branes. 
 
 Both these examples are  set in the context of Type  IIB  supergravity,  however some of the most exciting recent developments in holography have taken place in the different regime of M-theory, most notably the discovery of the ``class ${\cal S}$'' or $T_N$  ${\cal N}=2$ theories \cite{Gaiotto:2009we}, their ${\cal N}=1$ cousins \cite{Benini:2009mz} and their holographic duals \cite{Gaiotto:2009gz}. The lack of a conventional Lagrangian description makes the study of their holographic dual geometries of paramount importance for such theories.  One could wonder whether analogous flows triggered by VEVs or mass deformations arise in these theories and if they can be given a holographic description.   Indeed, the construction of  ${\cal N}=1$ theories from ${\cal N}=2$ counterparts by integrating out the adjoint scalars living in the ${\cal N}=2$ gauge multiplets of $T_N$ quivers given in  \cite{Benini:2009mz}, \cite{Bah:2012dg} suggests this is indeed possible although the geometrical description of this flow is not known.  
 
 The direct construction of holographic geometries encoding flows in the landscape of class ${\cal S}$ theories is a rather hard problem and to date there are no clear examples and few clues.  In this work, we will take a first step in this direction by giving some solutions of M-theory that have many of the properties we expect from these flows.  The way we shall arrive at these solutions is by harnessing a certain transformation of supergravity solutions known as non-Abelian T-duality.  

The non-Abelian T-duality procedure generalises the regular notion of T-duality to the context of non-Abelian isometry groups
\cite{delaossa:1992vc,Giveon:1993ai,Alvarez:1993qi}.   
At the level of supergravity it is anticipated--and though well checked\footnote{See \cite{Itsios:2012dc,Jeong:2013jfc,Kelekci:2014ima} for discussion on the subset of solutions where the solution generating nature of the duality is established.}, not yet completely established--that this is a solution generating transformation.   At the level of the world-sheet, unlike regular T-duality this should not be viewed as an exact equivalence of  CFT's since it is not expected to hold at all orders in string genus perturbation theory.  Indeed, 
 it was suggested in \cite{Giveon:1993ai}, 
that non-Abelian T-duality is a transformation
 between two {\it different} world-sheet CFTs.
In the context of the large N limit of holography 
(also with $g_s\to 0$ and $\alpha'\to 0$)  contributions associated with the genus expansion are suppressed and we might expect this transformation to have utility.   
 We will apply this non-Abelian T-duality  procedure to both the KM and the KW flows in type IIB and use it to produce what we believe will be rather 
prototypical supergravity solutions in M-theory that can describe holographic flows.  

The reader may wonder why one should resort to the more complex {\em non-Abelian} T-duality to achieve this goal; the reason is that performing regular (Abelian) T-dualities in conifold type backgrounds  either breaks supersymmetry (e.g. dualising around the $U(1)_R$ direction) or creates singularities (e.g. dualising  on a shrinking cycle).  The geometries we find will have ${\cal N}=1$ 
supersymmetry along the flow with enhancement in either the UV or IR. 

In \cite{Sfetsos:2010uq} it was shown that when a non-Abelian 
T-duality is applied to an $SU(2)$ subgroup of isometries 
of the five-sphere in $AdS_5 \times S^5$ 
(or its $\mathbb{Z}_2$ quotient) what results 
is a target space geometry that is in many ways 
rather similar to the holographic duals of the ${\cal N}=2$ 
theories presented in  \cite{Gaiotto:2009we}.  
We will elaborate further on this and the field theory interpretation in what follows.  In a similar fashion, the non-Abelian T-dual of an $SU(2)_L$ subgroup of $AdS_5 \times T^{1,1}$ performed in  \cite{Itsios:2012zv,Itsios:2013wd} gives rise to a geometry that shares many features with that corresponding to the ${\cal N}=1$ 
theories given of   \cite{Bah:2012dg}.   
Given these connections it is natural to  expect, 
that performing similar non-Abelian T-dualisation of flows in IIB is a good starting point with which to find flows in the M-theory setting of ``class ${\cal S}$''.  Although there is clearly more to be understood about the field theories corresponding to these examples of flows we present here, we are hopeful that knowledge of the geometries contained in this work will prove a helpful stepping stone  towards the broader question of finding more general holographic flows in M-theory.

 The structure of this manuscript is as follows:  
 
 In Section 
\ref{sec:NABTintro} we provide an introduction to the central tool of  non-Abelian T-duality (NATD). In Section \ref{sec:NABThol} we will discuss the non-Abelian T-dual of $AdS_5 \times S^5$ and further elaborate on its holographic dual description.  We then turn to flowing geometries by first presenting in Section \ref{sec:Ansatz} a wide ansatz for flows within the $AdS_5/CFT_4$ correspondence in type IIB supergravity. This ansatz incorporates both the KW and the KM flow.   In Section \ref{sec:Tdualansatz} we present the non-Abelian T-dual of this ansatz and explicitly demonstrate, using the technology of pure-spinors, that it satisfies the equations of motion and Bianchi identities of IIA supergravity.  
 
 In  Section \ref{sec:KM}, we  focus our attention  on the KM flow where the known analytic form of the background allows a more detailed study.  We begin the section with a brief recap of the KM flow  reviewing mostly known material but take the opportunity to clarify some points concerning brane charges, supersymmetry and central charges of the original solution that have not been discussed in the existing bibliography.   In Section  \ref{sec:KMdualprobes}, we present new result for the non-abelian T-dual of this geometry and we investigate a number of holographic observables in this geometry including the corresponding Baryonic condensate and an axionic string.    We close the paper with some conclusions and some useful technical appendices.
 
 \section{ An introduction to non-Abelian T-duality} \label{sec:NABTintro} 
 
 Since it may be less familiar to the reader let us begin by introducing the main technical tool of our work: non-Abelian T-duality.   Non-Abelian T-duality is the natural
 extension of T-duality of $U(1)$ isometries to the case of non-Abelian isometry groups in target space and goes back to the pioneering work 
\cite{delaossa:1992vc,Giveon:1993ai,Alvarez:1993qi}.  More recently, following the first implementation of non-Abelian T-duality in Ramond-Ramond backgrounds of supergravity \cite{Sfetsos:2010uq}, this has been actively exploited     \cite{Itsios:2012zv}-\cite{Araujo:2015dba}
 as a solution generating tool of supergravity 
particularly in the context of the $AdS/CFT$ correspondence.

 
The essential idea of non-Abelian T-duality is very similar in spirit to the familiar Abelian T-duality. One takes a 2d string $\sigma$-model on a target space and performs a Buscher 
  \cite{Buscher:1987qj,Buscher:1987sk, Rocek:1991ps}   dualisation procedure along the directions of a target space isometry group.  That is, in the 2d $\sigma$-model one introduces some gauge fields to gauge the global transformations associated to the isometries. To avoid extra degrees of freedom into the $\sigma$ model (at least classically), a Lagrange multiplier term is added to enforce that the field strength of the gauge fields vanishes.  Integrating out the Lagrange multipliers returns one to the initial $\sigma$-model.  On the other hand, integrating out the gauge fields whilst retaining the Lagrange multipliers provides a dual theory which after gauge fixing can be re-expressed again as a $\sigma$-model but in a different  T-dual target space.  
 
 Let us illustrate this with the simplest example; the principal chiral model on a group manifold $G$.  As a non-linear sigma model this has an action (in world sheet light cone coordinates $\sigma^\pm = \frac{1}{2} (\tau \pm \sigma)$)
\beq\label{eq:SPCM}
S_{PCM}= \frac{-\kappa^2}{2\pi}\int d^2 \sigma\, \Tr (g^{-1} \partial_- g  g^{-1} \partial_+ g)=   \frac{ \kappa^2}{2\pi} \int d^2\, \sigma \delta_{ij} L^i_\mu  L^j_\nu \partial_- X^\mu \partial_+ X^\mu   \ , 
\eeq
in which we have introduced a  group element $g\in G$ parametrised by local coordinates $X^\mu$, $\mu = 1 \dots \dim G$ and Maurer-Cartan forms $L^i = - i \Tr (g^{-1}dgT^i)$    satisfying $dL^i = \frac{1}{2} f^{i}{}_{jk} L^j \wedge L^k$ and generators of the algebra  $[T_i ,T_j] =  i f_{ij}{}^k T_k$ normalised such that $Tr(T_i T_j)=\delta_{ij}$.   For the case of $G= SU(2)$ the target space is just the round $S^3$ with metric\footnote{\label{foot:Euler}In Euler angles $g = e^{\frac{i}{2} \tilde{\phi}\tau_3 }e^{\frac{i}{2}  \tilde{\theta}\tau_2 }e^{\frac{i}{2} \tilde{\psi}\tau_3} $ the $SU(2)$ left invariant forms are 
\beq
\sqrt{2} L_1 = -\sin \tilde{\psi} d\tilde\theta + \cos\tilde{\psi} \sin\tilde\theta d \tilde\phi \ , \quad \sqrt{2} L_2 =  \cos \tilde{\psi} d\tilde\theta + \sin \tilde{\psi} \sin\tilde\theta d \tilde\phi \ , \quad \sqrt{2}L_3= d\tilde\psi + \cos\tilde{\theta} d\tilde{\phi} \ . 
\eeq
 }
\beq
ds^2= \lambda^2(L_1^2 + L_2^2  +L_3^2) \ , 
\eeq
where $\lambda^2$ is related to the dimensionless coupling $\kappa$ via $\kappa^2 = \frac{\lambda^2}{\alpha^\prime}$ and the scalar curvature of this space is $R= \frac{3}{\lambda^2}$. 

The target space has a $G_L\times G_R$ isometry and we will dualise the $G_L$ action.  In the $\sigma$-model eq.~\eqref{eq:SPCM} we introduce gauge fields $A=i A_i T_i$ to promote derivates to covariant derivatives $\partial_\pm \rightarrow D_\pm = \partial_\pm - A_\pm$ and supplement the action with a Lagrange multiplier term, 
\beq
S_{Lag} = \frac{\mu i }{ 2\pi} \int d^2 \sigma\,  \Tr v F_{+-} = \frac{\mu}{2 \pi} \int d^2 \sigma\,   v_i\partial_- A_+^i    - v_i  \partial_+ A_-^i - A_+^i f_{ij}{}^k v_k A_-^j
\eeq
  where the field strength $F_{+-} = \partial_+ A_- - \partial_- A_+ - [A_+ , A_-]$.  Evidently $\mu$ can be absorbed by scaling of $v$ but is useful to keep track of.

One now performs an integration by parts on the Lagrange multiplier term so that the gauge fields enter the action algebraically without derivatives and can be integrated out.  The next step is to fix the $G_L$ symmetry; here there may be several options but in this paper we will choose the simplest which is to set $g=\mathbb{I}$ such that the Lagrange multipliers play the r\^ole of coordinates in the dual $\sigma$-model.  In this gauge, the equations of motion for the components of the gauge field read 
\beq \label{eq:Apmrule}
A_+   = - \mu M^{-1} \partial_+v  \ , \quad A_-   = \mu M^{-T} \partial_-v \ , \quad M_{ij} = \kappa^2 \delta_{ij}+ \mu f_{ij}{}^k v_k  \ .
\eeq
Eliminating the gauge fields with these equations gives the dual action to eq.~\eqref{eq:SPCM}
\beq
\widehat{S} = \frac{\mu^2}{2\pi} \int d^2 \sigma \, \partial_- v^T  M^{-1}  \partial_+v  \ . 
\eeq
Notice that the T-dual model has a metric with rather complicated coordinate dependence since the $v^i$ enter explicitly into the definition of  $M_{ij}$.
The T-dual geometry is
\be 
\widehat{ds^2}= \frac{\mu^2 \alpha^\prime}{2}(M^{-1}+ M^{-T})^{ij}dv_i dv_j \ , \quad \widehat{B_2}= \frac{\mu^2 \alpha^\prime}{4}(M^{-1}- M^{-T})^{ij}dv_i \wedge dv_j \ , \quad \widehat{\Phi}= -\frac{1}{2}\log \det\left(\mu^{-2} M\right) \ ,  
\ee
in which we note the dilaton contribution that arises from performing this procedure in a path integral. For the case of $G_L= SU(2)$ the  geometry associated to the round $S^3$ reads 
\beq\label{eq:SU2dual}
\widehat{ds^2} = \alpha^\prime\frac{2}{ \k^2} d\rho^2 +\frac{  \alpha^\prime   \rho^2\frac{\kappa^2}{2}}{ (\frac{\kappa^4}{4}+\rho^2)} ds^2(S^2)\ , \quad \widehat{B_2} = \frac{\alpha^\prime \rho^3}{(\frac{\kappa^4}{4}+\rho^2)} \textrm{vol}(S^2) \ ,  \quad \widehat{\Phi} = - \frac{1}{2} \log (\frac{\kappa^2}{2} ( \frac{\kappa^4}{4}+ \rho^2)) \ , 
\eeq in which we have transformed the $v^i$ Lagrange multipliers into spherical coordinates and fixed $\mu = \sqrt{2}$  for later convenience.  The presence of a two-sphere in eq.~\eqref{eq:SU2dual} reflects a residual $SU(2)_R$ symmetry that was untouched by the dualisation\footnote{An interesting feature is that if one takes the limit $\rho\rightarrow \infty$ in \eqref{eq:SU2dual} then the NS sector matches that obtained by performing an abelian T-duality along the one of the Euler angles (precisely the $U(1)$ associated to $\tilde\psi$ defined in Footnote \ref{foot:Euler}). To complete such an identification one must rescale the dilaton and  make an identification of $\rho$ in this limit with a periodic variable.  In this sense one finds a limit in which non-Abelian T-duality abelianizes;  the idea of such a relation was suggested to  us by Jos\'e Luis Barb\'on.}. In general however, any symmetries that do not commute with the dualised isometry group will be destroyed.

In the present context, we wish to perform such a dualisation procedure in a background supported by RR flux.  This is rather more delicate since one needs to use an appropriate string theory formulation that incorporates the RR background fields. However, it was proposed in \cite{Sfetsos:2010uq}
 that the transformation rules of the RR sector can be deduced using just knowledge of the NS-sector.   The  crucial point is that left and right movers on the world sheet have different transformation properties under the duality---viz. eq.~\eqref{eq:Apmrule}.  After duality   one finds left and right movers couple to different sets of frame fields, call them $\hat{e}_+$ and $\hat{e}_-$ but since these frames define the same T-dual geometry they must be related by a local Lorentz transformation $\hat{e}_+^i = \Lambda^i{}_j \hat{e}_-^j$. This transformation induces an action $\Omega$ on spinors defined through the properties of $\gamma$-matrices  $\Omega^{-1} \Gamma^i \Omega =  \Lambda^i{}_j  \Gamma^j$. Space time spinors will be transformed under T-duality by this matrix $\Omega$.   The dual RR fluxes are then given by acting with this $\Omega$ matrix, 
 \beq\label{eq:RRrule}
e^{\widehat{\Phi}} \widehat{\slashed{\bf{ F}}}= e^{\Phi} \slashed{\bf{ F}} \cdot \Omega,
 \eeq
 where we consider the formal sum of forms\footnote{We work in the democratic formalism in which all degrees of fluxes are considered and Hodge duality is implemented afterwards \cite{Bergshoeff:2001pv}.}
 \beq
{\bf{ F}} =  \sum_{n=0}^5 F_{2n} \quad {\textrm{in Type IIA}}  \ , \quad {\bf{ F}} = \sum_{n=0}^4 F_{2n+1}    \quad {\textrm{in Type IIB}}   \ , 
 \eeq 
 and the slashed notation indicates contraction with $\Gamma$-matrices.\footnote{In the limit of Abelian T-duality of a single direction, call it $\theta$, T-duality acts as a parity on left movers and the corresponding $\Omega$ matrix will simply be $\Gamma^\theta$.  Then the T-duality rule  eq.~\eqref{eq:RRrule} boils down to erasing legs of flux that wrap the dualised circle and adding them when they don't, thereby replicating the action of T-duality on D-branes.}  This transformation rule can also be understood as a generalisation of the Fourier-Mukai transformations \cite{Gevorgyan:2013xka} and also gives rise to an action on supersymmetries \cite{Sfetsos:2010uq,Barranco:2013fza,Kelekci:2014ima}.

 Whilst here we outlined the procedure for the simplest example of a round metric on a group space, one set up a similar dualisation for a more general scenario in which the space admitting the $SU(2)$ isometry can be fibered non-trivially over other ``spectator'' directions.  A comprehensive treatment of this duality including spectator direction can be found e.g. in appendices of \cite{Itsios:2013wd}. Also,  in Section \ref{sec:Ansatz} of this paper we provide a set of ``Buscher rules'' at least for a wide class of geometries including our present interests.
 In the following, we will apply the formalism of this section to the example of $AdS_5\times S^5$. While this is not new in the bibliography, 
we will discuss new features, leading to a sharper 
dual field theoretical understanding of the resulting geometry.
   \section{Comments on the non-Abelian T dual of $AdS_5\times S^5$ }
\label{sec:NABThol}
Before moving to flowing geometries let us first look at   
$AdS_5\times S^5$  and comment on the relation between its 
non-Abelian T-dual,   first worked out in 
\cite{Sfetsos:2010uq}, and  
geometries dual to
Gaiotto QFTs \cite{Gaiotto:2009gz}.
 
 We start  with a background of the form,
\bea\label{eq: Ads5s5}
& & ds^2=\frac{ 4 R^2}{L^2}dx_{1,3}^2 +\frac{ 4 L^2}{R^2}dR^2 + L^2 \Big[4 d\alpha^2
+4 \sin^2\alpha d\beta^2 + 2 \cos^2\alpha (L_1^2+L_2^2+L_3^2) \Big],\nonumber\\
& & F_5= (1+ \star) \frac{64 }{g_sL^4} R^3 dR \wedge  d^4 x  \label{ads5xs5}, 
\eea
where $4 L^4= \pi g_s N \alpha'^2$ and $L_i$ are left 
invariant forms of $SU(2)$ of the previous section, normalised as in 
Footnote 2. We set
$g_s=1$ in the rest of this section. 

After non-Abelian T-duality on the $SU(2)$ parametrised by the $L_i$,  using eq.~\eqref{eq:SU2dual} with $\kappa = \sqrt{2/\alpha^\prime}  L\cos \a$ 
we arrive to a background of the form
\bea
& & \widehat{ds^2}=\frac{4R^2}{L^2}dx_{1,3}^2 +\frac{4 L^2}{R^2}dR^2 
+ L^2 \Big[4d\alpha^2
+4\sin^2\alpha d\beta^2 \Big]+\frac{\alpha'^2}{L^2\cos^2\alpha} 
d\rho^2 +\nonumber\\
& & 
\qquad\qquad \frac{\alpha'^2L^2\cos^2\alpha \rho^2}{\alpha'^2 \rho^2 
+ L^4\cos^4\alpha}(d\chi^2 +\sin^2\chi d\xi^2).\nonumber\\
& & \widehat{B}_2=\frac{\alpha'^3 \rho^3}{\alpha'^2 \rho^2 +L^4 \cos^4\alpha}
\sin\chi d\chi \wedge d\xi;\;\;\; 
e^{-2\widehat{\Phi}}=\frac{ L^2 \cos^2\alpha}{\alpha'^3}
(L^4 \cos^4\alpha +\alpha'^2 \rho^2).\nonumber\\
& & \widehat{F}_2=\frac{8L^4    }{\alpha'^{3/2}}
\sin\alpha \cos^3\alpha d\alpha \wedge d\beta,\;\;  \hat{A}_1=-\frac{2L^4   }{\alpha'^{3/2}} \cos^4\alpha d\beta. \;\; \widehat{F}_4= \widehat{B}_2\wedge \widehat{F}_2.
\label{ads5xs5natd}
\eea
  
In order to have quantised charges $Q_{Page,D6}=N_{D6}$ and $Q_{Page, D4}=0$, we have $L^4= \frac{ N_{D6}}{2}\alpha'^2$.
Note that we could have kept the relation $4 L^4 =\pi N {\alpha'}^2$  obtained before the duality,  but this would necessitate a constant rescaling of the RR sector in order to have well quantised charges and a corresponding shift in the  dilaton to solve the EOMs.  Notice that  at    
$2\alpha=\pi$, the background above is singular.

We now make contact with the backgrounds presented by Gaiotto and Maldacena
\cite{Gaiotto:2009gz} and developed in the papers 
\cite{ReidEdwards:2010qs}, \cite{Aharony:2012tz}. 
We will first take out a global $\alpha'$-factor in the metric which we will do with the coordinate transformations used in \cite{Sfetsos:2010uq},
\beq
\rho=2\frac{L^2}{\alpha'} \eta\ ,\;\;\;\; \sin\alpha=\sigma \ ,\;\;\;\;   u=\frac{R}{\alpha'}
\label{cambio}
\eeq
 in which we introduce the  usual
`energy' coordinate $u$.  We will 
find that the background in eq.(\ref{ads5xs5natd}) consists of 
a five dimensional Anti-de Sitter space of radius $\mu^2=\frac{L^2}{\alpha'}$
times a manifold $\Sigma_5$, together with rescaled NS and RR fields,

\bea
& & \frac{\widehat{ds}^2}{\alpha'}=4 \frac{u^2}{\mu^2}dx_{1,3}^2 +
4 \frac{\mu^2}{u^2}du^2 
+ \mu^2 \Big[4\frac{d\sigma^2}{1-\sigma^2}
+4\eta^2 d\beta^2 +\frac{4}{(1-\sigma^2)} 
d\eta^2 \Big]+\nonumber\\
& & \qquad \qquad
\frac{4\mu^2 \eta^2 (1-\sigma^2)}{4\eta^2 
+ (1-\sigma^2)^2}(d\chi^2 +\sin^2\chi d\xi^2).\nonumber\\
& & \widehat{B}_{2}=\frac{8\alpha'\mu^2 \eta^3}{4 \eta^2 + (1-\sigma^2)^2}
\sin\chi d\chi \wedge d\xi;\;\;\; 
e^{-2\widehat{\Phi}}=\mu^6(1-\sigma^2)
[ (1-\sigma^2)^2+4 \eta^2].\nonumber\\
& & \widehat{A}_1
= -2\mu^4\alpha'^{1/2} (1-\sigma^2)^2 d\beta. 
\label{ads5xs5natd3}
\eea
Let us now consider a generic Gaiotto-Maldacena 
background in type IIA \cite{Gaiotto:2009gz}. It reads,
\bea
& & ds_{IIA,st}^2=\alpha'(\frac{2\dot{V} -\dot{\dot {V}}}{V''})^{1/2}
\Big[  4 AdS_5 +\mu^2\frac{2V'' \dot{V}}{\Delta} 
{d \Omega^{2}_2(\chi,\xi)}+\mu^2\frac{2V''}{\dot{V}}  
(d\sigma^2+d\eta^2)+ \mu^2\frac{4V'' \sigma^2}{2\dot{V}-\dot{\dot{V}}} 
d{\beta}^2 \Big], \nonumber\\
& & A_1=2\mu^4\sqrt{\alpha'}
\frac{2 \dot{V} \dot{V'}}{2\dot{V}-\dot{\dot{V}}}d{\beta},\;\;\;\; 
e^{4\Phi}= 4\frac{(2\dot{V}-\dot{\dot{V}})^3}{\mu^{12}V'' \dot{V}^2 \Delta^2}, 
\quad {\Delta = (2 \dot{V} - \dot{\dot{V}}) V'' + (\dot{V}')^2} \ ,  \nonumber \\
& & B_2=2\mu^2\alpha' (\frac{\dot{V} \dot{V'}}{\Delta} -\eta) 
d\Omega_2,\;\;\; {A}_3={-} 4\mu^6 \alpha'^{3/2}
\frac{\dot{V}^2 V''}{\Delta}d{\beta} \wedge d\Omega_2.
\label{metrica}
\eea
 in which we defined a potential $V= V[\sigma,\eta]$ 
and its derivatives $V'=\partial_{\eta} V$ and $\dot{V}=\sigma\partial_{\sigma}V$. 
The  two-sphere $d \Omega^{2}_2(\chi,\xi)$ is parametrised by 
the angles $\xi$ and $\chi$ with corresponding volume 
form $d\Omega_{2}= \sin\chi d\xi \wedge d\chi$. The usual definition
$F_4= dC_3 + A_1\wedge H$ was also used.

Comparing both IIA configurations in eqs.(\ref{ads5xs5natd3}) and (\ref{metrica}), 
one can show that the background in eq.(\ref{ads5xs5natd}) is
 of the form of those
written by Gaiotto and Maldacena  
\cite{Gaiotto:2009gz}. This is not very surprising, since these 
solutions with an $AdS_5$ factor and preserving 
${\cal N}=2$ SUSY have been classified in 
\cite{Lin:2004nb}, \cite{OColgain:2010ev}.  
The problem of writing  IIA/M-theory  solutions boils 
to  finding the function $V(\sigma,\eta)$, which in turn 
reduces to resolving an electrostatic problem---
a Laplace equation for the function $V(\sigma,\eta)$ with a given charge density $\lambda(\eta)$,
\beq
\partial_\sigma[\sigma \partial_\sigma V]+\sigma \partial^2_\eta V=0,\;\;\;\;\;\;\;\lambda(\eta)= \sigma\partial_\sigma V(\sigma,\eta)|_{\sigma=0}.\label{ecuagm}
\eeq
What is actually interesting is to find the potential function
for the solution in eq.(\ref{ads5xs5natd3}).
This was done by Sfetsos and Thompson in \cite{Sfetsos:2010uq}, the result is
\beq
V_{ST}=\eta(\log\sigma -\frac{\sigma^2}{2})+\frac{\eta^3}{3}.
\eeq 
There is a relation between $V_{ST}$ and  
the potential function $V_{MN}$, characterising the solutions in
\cite{Maldacena:2000mw}. This potential reads---for a single $M_5$-brane,
\bea
& & 2V_{MN}(\sigma,\eta)=\sqrt{\sigma^2+(1+\eta)^2}-\sqrt{\sigma^2 +
(1-\eta)^2}+\\
& & +(1-\eta)\log[\frac{1-\eta}{\sigma} +\sqrt{1+(\frac{1-\eta}{\sigma})^2}]-
(1+\eta)\log[\frac{1+\eta}{\sigma} +\sqrt{1+(\frac{1+\eta}{\sigma})^2}].
\nonumber
\eea
This solution is interesting because, as it was shown in
\cite{ReidEdwards:2010qs}, \cite{Aharony:2012tz}, 
a very general $V(\sigma,\eta)$ solving eq.(\ref{ecuagm}),
can be written as a linear combination of
$V_{MN}$ for different number of $M_5$-branes.
Let us expand the function  $V_{MN}$ above in powers of $\eta^m \sigma^n$
such that $m+n<5$. We obtain,
\beq
V_{MN}\sim V_{app}= \frac{\eta^3}{6} +\eta\Big(\log(\frac{\sigma}{2}) -\frac{\sigma^2}{4} \Big)+....
\label{vapp}
\eeq
This approximate potential satisfies the differential equation in (\ref{ecuagm}). Both $V_{app}$ and $V_{ST}$ give the same density 
of charge $\lambda(\eta)=\eta$.
It is interesting to notice that adding more terms to the expansion,
one does not obtain a solution. In this sense, this is vaguely reminiscent of a 
Penrose limit---see the paper 
\cite{Polychronakos:2010hd}
for similar ideas expressed in a  different context. 

The careful reader observed that between 
$V_{ST}$ and $V_{app}$ there are some discrepancies 
in numerical factors.  Nevertheless, both these potentials
give place to the same background after a rescaling of the coordinates and 
Newton constant,
as we show in Appendix \ref{appendixpotentials}. 
We can lift the configuration to
a solution of eleven-dimensional supergravity, 
using that $\kappa^{2/3}=(\frac{\pi}{2}L_P^3)^{2/3}\sim \alpha'$, see Appendix \ref{appendixpotentials} for the details.

In summary, the non-Abelian T-duality of $AdS_5\times S^5$ is {\it generating} 
a background of the Gaiotto-Maldacena type, characterised by a function $V_{ST} (\sigma,\eta)$, a solution to a very involved
partial differential equation. This generated solution captures the small
 region close to the point $(\eta,\sigma)=(0,0)$ of a more generic background found in \cite{Maldacena:2000mw}.
The generating technique might suggest the  ansatz for the more generic
backgrounds.

Let us study now some interesting QFT observables 
as read from  the geometry in eq.(\ref{ads5xs5natd}).
\subsection{Central Charge and Page Charges}
We start by revisiting the treatment of Page charges
developed in \cite{Macpherson:2014eza}. There, it was shown that
\bea
& & Q_{Page D4}= 0,\;\;\; Q_{Page, D6}= N_{D6},
\eea
where normalisations in eq.(\ref{ads5xs5natd}) have been chosen to 
have proper quantisation.  Now we want to advance an interpretation of the $\rho$ coordinate---or $\eta$ after the change of variables 
in eq.~\eqref{cambio}-- which is that whilst $\rho$ is formally a non-compact variable, it is segmented in intervals of length $\pi$ by the presence of NS5 defects.  This interpretation builds on a subtle argument proposed in the papers  \cite{Lozano:2013oma,Lozano:2014ata}  and relies on two crucial points. First that the Page charges defined above are 
not invariant under large gauge transformations and second, 
that in the geometries we consider there is a periodic quantity 
$b_0$, defined as,
\beq
b_0=\frac{1}{4\pi^2\alpha'}\oint_{\Sigma_2} \widehat{B}_2  \subset  [0,1] \ .
\label{b0zz}
\eeq
In slightly different contexts, it was shown in \cite{Lozano:2013oma,Lozano:2014ata} that geometries produced by non-Abelian T-duality typically have such a two-cycle about which $b_{0}$ is defined and moreover that the expression obtained for $b_{0}$ depends on $\rho$.
Here the two-cycle is given by 
\beq
\Sigma_2=[\chi,\xi],\;\;\; \alpha=\frac{\pi}{2},\;\;\;\rho=fixed \ ,
\eeq
and one finds 
\beq
b_{0} = \frac{\rho}{\pi} \ . 
\eeq
To reconcile this result with the periodicity of $b_{0}$, one possibility could be that $\rho = \pi$ is a hard cut-off at the end off space. This seems  strange since the geometry is completely smooth at this point.  Instead we believe that the definition of $\hat{B}$ should actually be modified by a {\it piece-wise continuous large gauge transformation}  such upon moving from the interval $[0, \pi]$ to $[n\pi, (n+1)\pi]$  
\beq
\widehat{B}_2\to \widehat{B}_2-n\pi \alpha' \sin\chi d\chi \wedge d\xi,
\eeq
 thereby restoring the periodicity of $b_{0}$.   Notice, that the argument is not much different from the one that runs in determining the `period' of the R-symmetry coordinate
 $\psi$--reflecting the presence of the $U(1)_R$-anomaly--in dual to ${\cal N}=1$ QFT, see for example the papers \cite{Maldacena:2000yy}, \cite{Klebanov:2000hb}, \cite{Klebanov:2002gr}.  

Indeed, the point made in \cite{Macpherson:2014eza} was that
when we cross the boundaries $\rho=\pi, 2\pi,3\pi, 4\pi.... n\pi$
we need to perform a piece-wise continuous large gauge transformation of the $B_2$-field, that changes the Page charges as
\beq
\Delta Q_{D6}=0,\;\;\;\; \Delta Q_{D4}= - n N_{D6}.
\eeq
This suggests  that charge is `created' when we pass through $\rho=n\pi$ points. Hence, the charge of D4 branes is not
globally defined, but depends on the interval $[n\pi, (n+1)\pi]$ where we measure it. A very similar observation was made in the papers
\cite{Apruzzi:2013yva}, \cite{Rota:2015aoa}. In those papers they have a set of D8 sources and D6 branes whose Page charge is not globally defined, while here we
have NS-five branes playing the role of the D8's and  D4 branes with characteristics similar to their D6's. 

The expression 
of the $F_4, F_2$ fields indicates that we actually have a set of D4 and D6 crossed with NS-five 
branes extended along 
\beq
D4:[R^{1,3}, \rho], \;\;\;\; D6:[R^{1,3},\rho, \xi,\chi],\;\;\; NS5:[R^{1,3},\alpha,\beta]_{\rho=n\pi}.
\eeq
It is then possible to think, that we are in a situation where the D4 branes are blown-up into D6 branes on $S^2(\xi,\chi)$, 
due to the presence of the magnetic field $B_2$ via 
the Myers effect. Our Page charges indicate that the D4's are all blown into D6's in the interval $\rho~\epsilon~[0,\pi]$, but some remain in other intervals.
When lifting to M-theory, these D4's become M5 branes that extend on $AdS_5\times S^1$---matching our result with the discussion below eq.(4.8) in the paper \cite{Gaiotto:2009gz}. Additionally it can be shown, 
using the results of Section \ref{sec:Tdualansatz} below, 
that D4 and D6 branes are supersymmetric  at $\alpha
=\pi/2$.

Let us move to discuss 
a standing problem and propose a resolution for it. It relates to an issue with the central charge of the dual QFTs computed using the Type IIA background in eq.(\ref{ads5xs5natd}).

A discrepancy was observed 
in the previous bibliography related to the central
charge of the dual CFT. Indeed, while in the papers 
\cite{Maldacena:2000mw},
\cite{Gaiotto:2009gz} the central charge was found to scale as $c\sim N^3$---
a very unconventional scaling, with the cube of the number of M5 branes in the M-theory background--
in the papers 
\cite{Itsios:2013wd},
\cite{Macpherson:2014eza} the result in Type IIA  scales more conventionally 
like $c\sim N^2$--the square of the number of colour branes in the Type IIA configuration.
 Below we  analyse this in more detail, proposing an interpretation that makes compatible the QFT computation of \cite{Gaiotto:2009gz} with a type IIA calculation.

Before the discussion of central charges, we need to identify the number of NS-five branes.
The proposal in 
\cite{Bea:2015fja} is that the seemingly 
non-compact coordinate $\rho$ ---or $\eta$, after the change of variable in eq.(\ref{cambio})--should indeed be allowed
to vary in $[0,\infty]$. A large gauge transformation for the 
$B_2$-field has to be performed
every time we cross $\rho=n\pi$. The effect of this piece-wise continuous large gauge  transformation can be seen
if we calculate the flux of $\widehat{H}_3=d\widehat{B}_2$. Indeed, on the manifold $\Sigma_3=[\rho,\xi,\chi]$ with $\alpha=\frac{\pi}{2}$, we have that the NS field is
\beq
\widehat{H}_3|_{\Sigma_3}=\alpha' \sin\chi d\xi\wedge d\chi\wedge d\rho.
\eeq
Calculating one finds, 
\beq
\frac{1}{2\kappa_{10}^2 T_{NS5}}\int_{\Sigma_3}\widehat{H}_3= \frac{1}{2\kappa_{10}^2 T_{NS5}}\int_{0}^{(n+1)\pi} d\rho \int_{0}^{2\pi} d\xi \int_{0}^{\pi} d\chi \widehat{H}_3=(n+1).
\eeq
in which we have used that $2\kappa_{10}^2=(2\pi)^7 \alpha'^4$, $T_{NS5}=
\frac{1}{(2\pi)^5\alpha'^3}$. 
What brings the NS-five branes into existence is the piece-wise continuous character of the large gauge transformation.
This creation of a topological defect by a discontinuous transformation is also present in simple systems, see for example 
 the books \cite{Schwarz:1994cb}  (in electromagentism, the gauge potential corresponding to a solenoidal defect can be obtained from the vacuum gauge potential by performing a singular gauge transformation with a discontinuity or monodromy around a polar angle; the singular nature of the ``gauge transformation'' makes its presence known by giving a non-zero Wilson loop indicating a defect).

We can then identify that $N_{NS5}= (n+1)$ is the number of NS-five branes.
This enforces the picture advocated above---
and already advanced in \cite{Bea:2015fja}--
where there is one NS-five brane every time we cross the positions
$\rho=\pi,2\pi,3\pi,4\pi....$

Calculating the central charge also reinforces the picture above. Indeed, using the expressions relating central charges to internal volumes, see
\cite{Klebanov:2007ws} and for a generalization to include non-constant dilaton \cite{Macpherson:2014eza}
applicable in this case, one finds (normalizing such that $AdS_5\times S^5$ gives $c=\frac{1}{4} N_{D3}^2$, in agreement with \cite{Gubser:1998bc})  
\beq
c=\frac{1}{12} N_{D6}^2 N_{NS5}^3.\label{centralcharge}
\eeq
 This dependence with the number of NS five branes appears due to the integral 
$\int_{0}^{(n+1)\pi} d\rho \rho^2\sim (n+1)^3$. There is also the more canonical dependence with $N_{D6}^2$---that appears using the 
quantisation condition $2L^4= N_{D6}\alpha'^2$ discussed above-- and a numerical factor, both anticipated in 
the paper \cite{Macpherson:2014eza}. { In Appendix \ref{appendixpotentials}, 
we offer further evidence for the proposal made in this section}. 

The result of eq.(\ref{centralcharge}) above is, 
up to a normalisation factor,  the one  
obtained in equation (5.5) of the work by Gaiotto and 
Tomasiello \cite{Gaiotto:2014lca}.
In their case, the physical system is composed 
by $k=N_{D6}$ D6  branes,  D8 branes and $N=(n+1)$-NS five branes. 
The interpretation they propose for the QFT should apply to our case. 
We are probably dealing, after NATD, 
with a Gaiotto CFT  with $\Big((n+1) N_{D6}\Big)^3$ 
degrees of freedom that is orbifolded by
a $Z_{N_{D6}}$ group---see also \cite{Apruzzi:2015zna}
for a similar system in massive IIA. This matching supports 
our interpretation of the coordinate $\rho$, 
its range and  the large gauge transformation for the NS-field.

Having learnt 
something about the interplay between geometry and QFT, we close here this example of $AdS_5\times S^5$ and move to the core examples in this paper, namely the flows between conformal points. We will present first a generic form of flowing geometry, its non-Abelian T-dual and show that for these cases a NATD is indeed a solutions generating technique.

\section{A type  IIB ansatz for flows within the $AdS_5/CFT_4$ correspondence}
\label{sec:Ansatz}

We now turn our attention to holographic flow geometries and their non-Abelian T-duals.   Unlike Abelian T-duality, the presentation of the non-Abelian T-duality rules in complete generality is rather unwieldily and opaque without specifying any details of the seed geometry that is used as an input to dualisation.  One option (and this is what has been often adopted in the literature) is to fix a seed geometry completely, calculate its T-dual, and show that this is indeed a solution of supergravity.  This is somewhat unsatisfactory since conclusions are made on a case by case basis.   Instead here we will adopt an intermediate approach; we will specify an ansatz of IIB supergravity that is wide enough to incorporate many examples of interest but yet refined enough to lead to a tractable set of Buscher rules for the T-dual geometry\footnote{The most general ansatz non-Abelian T-dualised to date may be found in \cite{Kelekci:2014ima}.}.  Armed with such an ansatz and set of Buscher rules one can then simply specialise to the background of interest to investigate the details of its properties.  So our first step then is to specify an ansatz of IIB supergravity that captures both the KM and KW flows. 

Then, we consider performing a non-Abelian T-duality transformation on geometries of interest in the context of the $AdS_5/CFT_4$ correspondence. We require an $SU(2)$ isometry on which to dualise, 
so the topology of these solutions will be $R_{1,3}\times {\cal M}_3\times S^3$, where ${\cal M}_3$ will be non compact. We will assume that $F_1= F_3 = H_3=\Phi =0$ and take the ansatz 
  \beq\label{eq:ds10inXXX}
ds^2 =e^{2A}dx^2_{1,3}+ ds^2({\cal M}_3)+ \sum_{i=1}^3 \big(e^{i}\big)^2 \ , \quad F_5  =  (1+ \star){\cal F}_5 \ ,   \quad {\cal F}_5  =  {\cal F}_2 \wedge e^1 \wedge e^2 \wedge e^3 \ , 
\eeq
where $A$ has dependence on the coordinates of ${\cal M}_3$ and $e^{i}$ defines a vielbein on a squashed sphere which is fibered over  ${\cal M}_3$, namely
\beq\label{eq:ein}
e^i = \lambda_i (\omega_i + {\cal A}_i).
\eeq
Here  $\omega_i$ 
are a set of left invariant  Maurer-Cartan forms 
for the $SU(2)$, that
satisfy $d\omega_i =\frac{\epsilon_{ijk}}{2}  \omega_j \wedge \omega_k$,   ${\cal A}_i$ are one-forms on  ${\cal M}_3$ and $\l_i$ are functions on ${\cal M}_3$.  The Bianchi identity of $F_5$ requires that,
\beq\label{eq:BianchiF_5}
d( \l_1 \l_2 \l_3 {\cal F}_2) = 0 \ , \quad d(e^{4A} \star_3 {\cal F}_2)= 0 \ ,
\eeq
with ${\cal F}_2$  a two-form and $\star_3$, the Hodge dual, defined on  ${\cal M}_3$. In what follows it will also be useful 
to introduce a set of undetermined frame fields $h^i$ such that
\beq
ds^2({\cal M}_3)=\sum_{i=1}^3  (h^i)^2,\label{zaraza}
\eeq
where we also define $Vol({\cal M}_3)= h^1\wedge h^2\wedge h^3$.

At this point we make a restricting assumption that will nonetheless be sufficient for the solutions we consider in this work, as well as a good deal more of the literature. Let us assume that this background supports an $SU(3)$ structure on the 6d internal space specified by,
  \beq\label{eq: SU3str}
  J = h^3 \wedge e^3+ e^1\wedge e^2 +  h^1 \wedge h^2 \ , \quad  \Omega_h = (h^3 + i e^3)\wedge (e^1 + i e^2 ) \wedge (h^1+ i h^2) \ , 
  \eeq 
  with corresponding pure spinors \cite{Grana:2005sn}
  \beq\label{eq:ps}
  \Psi_+ = \frac{1}{8} e^{i \theta_+} e^{A} e^{-i J}  \ , \quad \Psi_- = -\frac{i}{8}  e^{i \th_-} e^A \Omega_h  \ , 
  \eeq
  which obey
  \beq
  d( e^{2 A} \Psi_- ) = 0 \ , \quad  d(e^{2 A} \Psi_+ ) - e^{2A} dA\wedge \bar\Psi_+ = - \frac{i}{8}e^{3A}\star_3 {\cal F}_2 \ ,
    \eeq
so that the solutions preserve (at least) $\mathcal{N}=1$ supersymmetry in 4d. Unpackaging these equations one finds the following set of independent constraints,
  \beq\label{eq:starF2}
  d(e^{2A} J) =0 \ , \quad d(e^{3A+ i \th_-} \Omega_h)=0   \  , \quad  \theta_+ =   \frac{\pi}{2}  \  , \quad   \star_3 {\cal F}_2 = 4 d A \ ,
  \eeq
  which impose certain conditions on $h^i,{\cal A}_i$.
  
 First let us consider  $d(e^{2A} J)=0$. One finds, from the components involving $\omega_1 \wedge \omega_3$ and $\omega_2 \wedge \omega_3$ that,
  \be\label{eq:Asol}
  {\cal A}_1 = {\cal A}_2 = 0 \ , 
  \ee 
  and from those involving $\omega_1 \wedge \omega_2$ we determine,
  \beq\label{eq:ersol}
  e^{2 A} \l_3 h^3 =d( e^{2A} \l_1 \l_2) \ . 
  \eeq
  The remaining components of  $d(e^{2A} J)=0$ imply
  \beq\label{eq:dthphi2}
  d\left( e^{2A }h^1\wedge h^2  - e^{2A}\l_{3}{\cal A}_3 \wedge h^3  \right) = 0 \ . 
  \eeq
	
Using the equation $d(e^{3A+ i \th_-} \Omega_h)=0$ we find,
  \begin{align}\label{eq:dthetaphi}
dh^1&=h^1\wedge\big(d(3A+ \log (\l_2\l_3))-\frac{\l_1}{\l_2\l_3}h^3\big)+h^2\wedge\big(\frac{\l_1}{\l_2}{\cal A}_3-d\theta_-\big),\nn\\[2mm]
dh^2&=h^2\wedge\big(d(3A+ \log (\l_1\l_3))-\frac{\l_2}{\l_1\l_3}h^3\big)-h^1\wedge\big(\frac{\l_2}{\l_1}{\cal A}_3-d\theta_-\big),
  \end{align}
as well as
	\beq\label{eq: weird}
d\big( {\cal A}_3-i\l_3^{-1} h^3\big)\wedge (h^1+ i h^2)=0
	\eeq
and the compatibility constraints
    \bea\label{eq:como}
   (\l_1^2 - \l_2^2)h^3 \wedge h^1&=& \l_3h^1\wedge    \left (\l_2 d\l_1 - \l_1 d\l_2 \right) -  \l_3(\l_1^2 - \l_2^2) h^2 \wedge {\cal A}_3 \ , \nn \\ 
       (\l_1^2 - \l_2^2) h^3 \wedge h^2 &=& \l_3 h^2 \wedge \left(  \l_2 d\l_1 - \l_1 d\l_2 \right) + \l_3(\l_1^2 - \l_2^2) h^1 \wedge {\cal A}_3 \ . 
  \eea
There is actually a degree of redundancy between these equations above and eq. \eqref{eq: weird}.  Either $\l_1=\l_2$ in which case the constraints are trivially satisfied and give no further information or $\l_1\neq\l_2$  whence one can show that eq. \eqref{eq: weird} follows from 
differentiating the constraints and applying eq. \eqref{eq:dthetaphi}. However all quoted expressions will be useful in the next sections.

Any background solution that fits into the ansatz of eq. \eqref{eq:ds10inXXX}  and satisfies eqs. (\ref{eq:Asol})-(\ref{eq:como}) will preserve supersymmetry in the form of an $SU(3)$-structure given by eq. \eqref{eq: SU3str}.  This ansatz is sufficient to include, at least, the following backgrounds of direct interest to the present paper:
\begin{itemize}
\item   $AdS_5 \times T^{1,1}$ with the $\omega_i$ corresponding to either the $SU(2)_L$ isometry {\em or} to the diagonal $SU(2)_{diag}$ isometry;  
 \item  $AdS_5 \times Y^{p,q}$ of  \cite{Gauntlett:2004hh,Martelli:2004wu} which has a unique $SU(2)$ isometry in the internal space;
\item The KM flow given by \cite{Klebanov:2007us};
\item The KW flow ansatz given by \cite{Halmagyi:2004jy};
\item  $AdS_5 \times S^5$ (of course  this only makes manifest ${\cal N}=1$ supersymmetry and the background preserves $\mathcal{N}=2$   after dualisation \cite{Sfetsos:2010uq}).
\end{itemize}

In appendix  \ref{sec:backgrounds} we provide a precise map of how the geometries listed above fall within this ansatz.     

\section{Type IIA/M-theory non-Abelian T-dual backgrounds of  
the flow ansatz}\label{sec:Tdualansatz}
In this section we present the non-Abelian T-dual of the ansatz of the previous section and give the $SU(2)$-structure it supports. This, combined with the vanishing of the Bianchi identities, which we also show,  provides a proof that the T-dual of this ansatz is always a solution of type IIA supergravity. That these are sufficient conditions was  spelled out in  \cite{Lust:2004ig,Gauntlett:2005ww}. In what follows we work with $\alpha'=1$ for simplicity.\footnote{The $\alpha'$ factors can be put back by the following replacements $v_i\to \alpha' v_i$, $\hat F\to \frac{1}{\alpha'~\!^{3/2}} \hat F$ and $e^{\hat{\Phi}}\to \alpha'~\!^{3/2}e^{\hat{\Phi}}$, which leave the EOM invariant.}

Following Appendix \ref{sec:NABTrules}, we perform a NATD transformation on the solution of the ansatz of eqs.(\ref{eq:ds10inXXX})-(\ref{zaraza}). After setting ${\cal A}_1 = {\cal A}_2 = 0$ as required by supersymmetry prior to dualisation we get the NS sector
\begin{align}
d\widehat{s}^2 &= e^{2A} dx_{1,3}^2+ ds^2({\cal{M}}_3)+ \sum_{i=1}^2 \widehat{e}^i_{\pm} \ , \nn\\[2mm]
\widehat{B} &= \frac{1}{\Delta}\bigg(\l_3^2 v_3 v_idv_i\wedge {\cal A}_3+\l_1^2\l_2^2\l_3^2 dv_3\wedge {\cal A}_3+\frac{1}{2}\epsilon_{ijk}v_i\lambda_i^2 dv_j\wedge dv_k\bigg),\nn\\[2mm]
e^{-2\widehat{\Phi}}&= \Delta =\l_1^2\l_2^2\l_3^2+ \l_1^2 v_1^2+ \l_2^2 v_2^2+ \l_3^2 v_3^2 \  , 
\end{align}
where the frame fields $\widehat{e}^i_{\pm}$ are the natural ones that arise from the Buscher procedure and are given by eq. \eqref{eq:eidents}, however an explicit form of a much nicer set of  frame fields is given in eq.~\eqref{eq:dualframesneat} below. 
The RR fluxes are given by
\beq
\widehat{F}_2= \l_1 \l_2\l_3 {\cal F}_2,~~~~~~~~\widehat{F}_4= (\widehat{B}+{\cal A}_3\wedge dv_3)\wedge \widehat{F}_2,
\eeq
and their Hodge duals by
\begin{align}\label{eq:simplefluxes}
\widehat{F}_6&=-\star_{10} \widehat{F}_4=-e^{4A}Vol_4\wedge v_idv_i\wedge\star_3{\cal F}_2,\nn\\[2mm]
\widehat{F}_8&= \star_{10} \widehat{F}_2 =\widehat{B} \wedge \widehat{F}_6+ e^{4A}Vol_4\wedge dv_1\wedge dv_2\wedge dv_3\wedge \star_3{\cal F}_2.
\end{align}
Clearly $d\widehat{F}_2=0$ by virtue of the first condition of eq. \eqref{eq:BianchiF_5}, 
it then follows that $d\widehat{F}_4=\widehat{H}_3\wedge \widehat{F}_2$ because ${\cal A}_3\wedge {\cal F}_2$ is a top-form on ${\cal M}_{3}$ and thus closed, so the Bianchi identities are automatically satisfied. The equations of motion of the higher fluxes can be shown to be solved in a similar fashion using the second condition of eq. \eqref{eq:BianchiF_5}. 
 
The T-dual pure spinors are given by 
\begin{equation}\label{eq:dualps}
 \widehat{\Psi}_- = \frac{1}{8} e^{A}e^{i\hat{\theta}_-} e^{-i j} \wedge z  \ , \quad \widehat{\Psi}_+ = \frac{1}{8} e^{A} e^{i\hat{\theta}_+} e^{\frac{1}{2} z\wedge \bar z} \wedge \omega,
\end{equation} 
which may be generated directly via the $\Omega$ matrix of eq. \eqref{eq: OmegaMatrix} via the map given in \cite{Barranco:2013fza}
\beq
\widehat{\Psi}_{\pm}= \Psi_{\mp}\Omega^{-1}.
\eeq 
The result can be neatly expressed after a frame rotation in terms of the internal vielbeins
 \begin{equation}\label{eq:dualframesneat}
  \begin{aligned}
\tilde{h}^3 &=   -\frac{   \l_1 \l_2\l_3}{\sqrt{ \Delta} } h^3 +  \frac{1}{\sqrt{ \Delta}} v_i dv_i\ ,  \quad 
\tilde{h}^{1}  = h^1 \ , \quad \tilde{h}^{2}=h^2 \ ,  \\ 
\tilde{e}^1 &= \frac{\l_2 \l_3 }{\sqrt{\Delta}} (dv_1-  v_2 {\cal A}_3) + \frac{v_1 \l_1 }{\sqrt{\Delta}} h^3 \ ,  \\
\tilde{e}^2 &= \frac{\l_1 \l_3 }{\sqrt{\Delta}} (dv_2 +  v_1 {\cal A}_3) + \frac{v_2 \l_2 }{\sqrt{\Delta}} h^3  \ , \\
\tilde{e}^3 &= -\frac{ \l_1 \l_2 }{\sqrt{\Delta} } dv_3 -  \frac{v_3 \l_3 }{\sqrt{\Delta}} h^3 \ .  
\end{aligned}
\end{equation}
In terms of these the T-dual $SU(2)$ structure is given by 
\begin{equation}
 z= v+i w=\tilde{h}^3 + i \tilde{e}^3 \ , \quad j = \tilde{h}^1 \wedge \tilde{h}^2 + \tilde{e}^1 \wedge \tilde{e}^2  \ , \quad \omega = (\tilde{h}^1 +  i \tilde{h}^2)\wedge(\tilde{e}^1 + i \tilde{e}^2)  \ , \quad  \hat\theta_+ = \theta_- \ , \quad  \hat\theta_- = \theta_+ = \frac{\pi}{2}  .
\end{equation} 
It is then relatively simple to plug this into the conditions the structure must obey and see that supersymmetry is indeed preserved. We relegate the details 
of this computation to Appendix \ref{appendix:proof}.

The succinct expressions for the T-dual geometries, frame fields and $SU(2)$ structure in this section are an important technical result of this work and provide a unified presentation of previous results in non-Abelian T-duality. It should be noted though that a more general proof of the solution generating nature of the non-abelian T-duality exists.
	
It was conjectured in \cite{Sfetsos:2010uq} that a condition for the preservation of supersymmetry is the vanishing Kossmann-Lie derivative of the corresponding Killing-spinor along all the Killing vectors, $k_a^\mu\partial_\mu$, generating the $SU(2)$ isometry, {\it i.e.} one requires
\be
{\cal L}_{k_a}  \epsilon = k_a^\mu D_\mu \epsilon + \frac{1}{4} \nabla_\mu k_{a \nu} \Gamma^{\mu\nu}\epsilon = 0 \ , \quad  a = 1\dots 3 \ . 
\ee
This conjecture was explicitly verified in  \cite{Kelekci:2014ima} by examining the transformations of dilatino and gravitino supersymmetry variations under T-duality. The Bianchi identities were also shown to follow in the T-dual from those of the seed solution. It then follows, by the results of \cite{Grana:2005sn,Lust:2004ig,Gauntlett:2005ww} that any supersymmetric solution with a metric that can be decomposed as in eq \eqref{eq:ds10inXXX}, \eqref{eq:ein} and arbitrary fluxes that respect the $SU(2)$ isometry, will be mapped to a supersymetric solution of type II supergravity, provided the Kossmann derivative vanishes. This of course gives no a priori information about the dual G-structure, which is the main achievement of this section.

 It would be interesting to generalise the results above 
to a more general ansatz, finding the general G-structure preserved. We note that the requirement of a vanishing Kossmann derivative of the Killing-spinor is equivalent to the vanishing of the Lie derivative when acting on the corresponding pure spinors. This hints at how such a generalization could be achieved.

\subsection{Lift to M-theory}
Since these are solutions of IIA supergravity with no Romans' mass they can be lifted in the usual way directly to eleven dimensions with  a metric
\be
ds_{11}^2 = e^{-\frac{2}{3}\hat\Phi } \left(   e^{2A} dx_{1,3}^2+  (\tilde{e}^i)^2 + (\tilde{h}^i)^2\right)+ e^{\frac{4}{3} \Phi} \left( dz + C_1 \right)^2\ , 
\ee 
where $z$ denotes the M-theory circle and $C_1$ is a potential such that $dC_1  = \widehat{F}_2$.  The geometry is characterised by an $SU(3)$ structure on the seven dimensional internal space specified by a one-form $K^\prime$, a two-form $J^\prime$ and a three form $\Omega^\prime$ \cite{Kaste:2003zd} obtained by lifting the $SU(2)$ structure defined above,
\begin{equation}
K^\prime = w e^{-\frac{1}{3} \hat{\Phi}} \ , \quad J^\prime =   e^{-\frac{2}{3} \hat\Phi}j + e^{\frac{1}{3} \hat{\Phi}} v\wedge (dz +C_1) \ , \quad \Omega^\prime = e^{i \hat \theta_+} \omega \wedge \left(e^{-\hat{\Phi}} v+ i (dz +C_1) \right) \ . 
\end{equation}
Using the expressions above these can be directly seen to obey the required differential equations  \cite{Kaste:2003zd} (see also \cite{Gauntlett:2005ww}),  
\be
d(e^{2\d}K^\prime)=0 \ , \quad d(e^{3\d}\Omega^\prime)= 0 \ , \quad d(e^{4\delta}J^\prime)= \star_7 e^{4\delta} G_4 \ , \quad  d(e^{2 \delta} J^\prime \wedge J^\prime) =-2 G_4 \wedge K^\prime e^{2\delta} \ ,
\ee
with $G_4= \widehat{F}_4 +  d\widehat{B} \wedge  (dz +C_1)$ obeying a Bianchi identity $d G_4 = 0 $ and the warp factor $\delta = A- \frac{1}{3}\hat{\Phi}$. 

We will now study the Klebanov-Murugan background and its non-Abelian T-dual as a simple applications of the formalism developed above. See Appendix \ref{sec:backgrounds}
for a compendium of other possible applications.

 \section{The Klebanov-Murugan Flow}\label{sec:KM} 

Here we shall study the Klebanov-Murugan (KM) background \cite{Klebanov:2007us}. 
This provides an example in which the solution is known analytically  
along the flow.
In the dual  field theory,  the flow corresponds to a  vacuum in which
certain baryonic operators acquire a VEV. 
On the gravity side this is represented by
 deformations of $AdS_5$ backgrounds.

To be more precise, consider D3 branes at the tip 
of the conifold or, more generally, other singular Ricci flat Kahler asymptotically
conical manifold.
Vacua obtained by moving the D3 branes stack away from the 
singularity, by resolution or deformation of the conifold, correspond to
a  form of symmetry breaking in the dual QFT. In all these cases, at energies 
low enough,
the field theory describing the stack of branes
will be ${\cal N}=4$ Super-Yang-Mills.

Using four complex coordinates $w_i$ to describe the conifold by the constraint $\sum_{i=1}^{4} w_i^2=0$, the deformation of the conifold is described by the six-manifold satisfying $\sum_{i=1}^{4} w_i^2=\epsilon$. More important in this section will be the {\it resolution} of the conifold, best expressed in terms of the variables $a_i,b_j$
\beq
w_1=a_1b_1,\;\;\; w_2=a_2b_2,\;\;\; w_3= a_1b_2,\;\;\; w_4=a_2b_1.\nonumber
\eeq
To describe the resolved conifold, we need to saturate the equation,
\beq
|a_1|^2+|a_2|^2-|b_1|^2-|b_2|^2= a^2,\nonumber
\eeq
subject to the identification up to a phase,
\beq
a_i\sim e^{i\nu}a_i,\;\;\;\; b_j\sim e^{-i \nu}b_j.\nonumber
\eeq

The IIB solution for D3 branes at the tip of the resolved conifold reads,
\beq
ds^2= H^{-1/2} dx_{1,3}^2 + H^{1/2}\Big(dr^2 + r^2 ds^2(X_5)\Big).\nonumber
\eeq
We will interpret the breaking of symmetry 
as replacing $H(r)\sim c/r^4$ by a more general 
Green's function. In this section it will be 
the Green function found by Klebanov and Murugan in
\cite{Klebanov:2007us}, which describes
a deformation of the Klebanov-Witten quiver field theory, obtained by giving a VEV to  bifundamental fields, subject to 
only non-mesonic operators getting a  VEV. This corresponds to a motion in the Kahler moduli space on the geometry side. In the field theory we study in this section, there will be an infinite tower of operators developing a vacuum expectation value. The operator with smallest 
dimension is taken to be the order parameter for the symmetry breaking. 
In the present example, the operator has 
dimension two and will be written explicitly below.
In the following, we write the Klebanov-Murugan background 
\cite{Klebanov:2007us}. We will complement this with some new calculations and  comments not present in the bibliography. We will then generate a new background
in Type IIA, by applying non-Abelian T-duality to the 
original Type IIB one in \cite{Klebanov:2007us}.  Furthermore, we study
different aspects of the strongly coupled QFT associated with our Type IIA geometry.

\subsection{The Klebanov-Murugan geometry}
In this section we review the Klebanov-Murugan Type IIB background \cite{Klebanov:2007us}. 
The singular conifold is given by the cone over $T^{1,1}$ which is an homogenous space of topology $S^2\times S^3$.  At the tip of the cone the $S^2$ shrinks but however can be resolved to give a finite sized $S^2$ of radius $a$.  This resolved conifold is topologically $R^4 \times S^2$  and an explicit Calabi-Yau metric is given by  
\beq
\begin{aligned}\label{eq:KMmetric}
ds^2_6 &= \frac{dr^2}{\kappa(r)} +ds^2(\mathcal{M}^3) +  A_2(r)^2 \left( d\theta^2 + \sin^2\theta d\varphi^2 \right) ,  \\
ds^2(\mathcal{M}^3) &= A_1(r)^2\big(\omega_1^2+\omega_2^2\big) + A_3(r)^2\big(\omega_3+\cos\theta d\varphi\big)^2,
\end{aligned}
\eeq
in which the functions are 
\begin{align}\label{eq:alpha}
A_1(r)^2&=\frac{r^2}{6}, \quad 
A_2(r)^2 =\frac{r^2+6 a^2}{6},\quad
A_3(r)^2  = \frac{\kappa r^2}{9}, \quad 
\kappa(r) = \frac{r^2+9 a^2}{r^2+6 a^2} . 
\end{align}
In the metric  
$ds^2(\mathcal{M}^3)$, we  have singled out a particular three dimensional subspace which contains the $SU(2)$ on which 
we will perform the non-Abelian T-duality. 

One can now consider $D3$ branes at some position $\vec{r}$ in this resolved conifold which gives rise to a warped supergravity solution supported by RR five form flux and constant dilaton,  
\beq
\begin{aligned}
ds^2 &= \frac{1}{L^2\sqrt{H(\vec{r})}} dx_{1,3}^2+ L^2\sqrt{H(\vec{r})} ds^2_6 , ~~~~
F_5 &=\frac{1}{L^4}\big(1+\star  \big) Vol_4\wedge dH(\vec{r})^{-1}  \ , \quad 
e^{\Phi} =1,
\end{aligned}
\eeq
where we are choosing to extract a factor of $L$ from $H$ with respect to the definition of \cite{Klebanov:2007us}, and as elsewhere we set $g_s=1$.
The function $H(\vec{r})$ solves a Laplace equation on the resolved conifold with delta-function sources at the location of the branes.  Indeed, if the branes are placed at a fixed location of the $S^2(\theta,\varphi)$, {which for simplicity we take to be the north pole to have $\varphi$-independence}, the warp function will be of the form $H(r,\theta)$. This yields a smooth solution of supergravity provided that the Bianchi identity (or Maxwell equation) for $F_5$ is satisfied. 
Away from the branes, the differential equation for $H(r,\theta)$ reads,
\bea
(9 a^2+r^2)r  \frac{\partial^2H}{\partial r^2} + 
6r\left(\frac{\partial^2 H}{\partial\theta^2}+\cot\theta \frac{\partial H}
{\partial\theta}\right)+(27a^2+5r^2)\frac{\partial H}{\partial r} =0.
\label{eq:diffH}
\eea
If we assume that $H(r)$ is a function of just the radial direction in the cone, this gives the Pando-Zayas Tseytlin solution \cite{Pando Zayas:2000sq} with a singularity characteristic of smearing the branes around the  $S^2$. 

 More interestingly, the non-smeared solution to eq. (\ref{eq:diffH}), allowing a more general position for the branes, was constructed by 
Klebanov and Murugan \cite{Klebanov:2007us}, 
as a superposition of an infinite number of harmonics. It reads,
\beq\label{eq:HKM}
H_{KM}(r, \theta)=\sum_{l=0}^\infty
 (2l+1)H_{l}^A(r)P_l(\cos\theta) \ , 
\eeq
in which $P_l$ are the standard Legendre polynomials and $H^A_l(r)$ are a set normalised hypergeometric functions whose exact details we shall not need for the moment but can be found in equation (33) of the paper 
\cite{Klebanov:2007us}.   This infinite summation localises the branes and indeed the solution of Pando-Zayas and Tseytlin  \cite{Pando Zayas:2000sq}  corresponds to truncating to just the zero mode, 
\beq
\label{eq:HlPZT}
H_{l=0}^A= \frac{2}{9 a^2 r^2}- \frac{2}{81 a^4} \log\left(1+ \frac{9 a^2}{r^2}\right)
\eeq
The field theory interpretation of the flow  described by the background with warp factor $H$ in eq.(\ref{eq:HKM}), 
corresponds to giving a VEV to the dimension two operator 
\begin{eqnarray}
{\cal U} = \frac{1}{N}\Tr(|a_1|^2 + |a_2|^2 -|b_1|^2 -  |b_2|^2   ), \nonumber
\end{eqnarray}
in which the $a_i$ and $b_i$ are the bi-fundamental matter fields of the quiver gauge theory.  
This operator has protected dimension $\Delta=2$ because it is in the multiplet of the baryonic current. Its VEV also indicates a breaking of the baryonic symmetry.
Indeed, placing the branes at the north pole of the blown up two-sphere is achieved by giving VEVs   $a_1=a_2=b_2=0$ and $b_1= a \mathbb{I}$.  In addition to providing a VEV for ${\cal U}$, this also gives VEVs to an infinite tower of operators.  This can be read from the infinite summation of harmonics that contribute to the solution in eq.~\eqref{eq:HKM}.
Each harmonic entering in the sum is a normalisable mode with a specific fall-off behaviour that defines in the usual way the dimension of the operator acquiring a VEV.  It was  shown in \cite{Klebanov:2007us}, that the presence of a VEV for the field $b_1$, breaks the gauge symmetry $SU(N)\times SU(N)\to SU(N)$ and the global symmetry $SU(2)\times SU(2)\times U(1)_B\times U(1)_R$ down to
$SU(2)\times U(1)\times U(1)$. The last two $ U(1)$'s are diagonal combinations between the broken $SU(2), U(1)_B, U(1)_R$. Also, it can be seen that the bifundamental fields $a_1, a_2, b_2$ now transform under the adjoint of the unbroken $SU(N)$. The integration out of $b_1$ leaves us with a cubic super-potential indicating that our low energy field theory is ${\cal N}=4$ Super-Yang-Mills. 
Other similar flows in more involved field theories have been studied in \cite{Pini:2014bea}.

Let us now consider some physical  and geometrical aspects of this solution.

\subsection{Charges in the Klebanov-Murugan geometry}
Let us analyse the quantisation of the RR charge, using the compact manifold defined by $X^5=[\theta,\varphi,\omega_1,\omega_2,\omega_3]$, we impose
\beq
\int_{X^5} F_5= 2\kappa_{10}^2 T_3 N_{D3},
\eeq
where 
\beq
2\kappa_{10}^2 = (2 \pi)^{7}  \alpha^\prime{}^4, \;\;\;\;\;T_3=\frac{1}{(2\pi)^3  \alpha'^2}, \;\;\;\;\;g_s=1, \nonumber
\eeq
the previous condition quantises $L^4=\frac{27}{4} \pi  N_{D3} \alpha'^2$ with $N_{D3} \in \mathbb{N}$ the number of D3 branes sourcing the solution. 

In the full KM solution the $D3$ charge is computed as
\beq\label{eq:D3charge}
\frac{1}{2\kappa_{10}^2 T_3}\int_{X^5} 
F_5=\frac{1}{8} N_{D3}(9a^2+r^2)r^3\int_0^{\pi} 
\sin\theta \frac{\partial H}{\partial r} d\theta .
\eeq
For the warp factor given in eq.~\eqref{eq:HKM},  this implies computing,
\beq
\int_0^\pi \sin\theta \partial_r H (r,\theta)
=\sum_{l=0}^\infty
 (2l+1)\partial_r H_{l}^A(r)\int_{0}^\pi 
\sin\theta P_l(\cos\theta)d\theta
= \sum_{l=0}^\infty
 (2l+1)\partial_r H_{l}^A(r) 2 \delta_{l,0} ,\nonumber
\eeq 
where we used the identity
\beq
\int_{-1}^{1} P_l(x)dx
= \frac{\sin(l\pi)}{l\pi}
\frac{2}{l+1}= 2\delta_{l,0}.\nonumber
\eeq
Indicating that only the zero-mode contributes. Hence we can use $H_{A,l=0}$ as given by eq. \eqref{eq:HlPZT}  and upon taking the derivative explicitly we find,
\beq
\int_0^\pi \sin\theta 
\partial_r H_{KM}(r,\theta)=-\frac{8}{r^3(r^2+9a^2)}.
\eeq
Substituting this result into eq.~\eqref{eq:D3charge} confirms indeed that the D3 brane charge is consistent in the full KM geometry and is given by  $Q_{D3}=-N_{D3}$.  
Notice that the sign just amounts to a choice of an orientation (also reflected in the way we computed the ten-dimensional dual $*_{10}$).
It is interesting to observe that the calculation above shows that the charge is all carried by the zero-mode, the Pando-Zayas-Tseytlin solution. The higher harmonics do not contribute to the quantised charge. 
\subsection{Limits of the Klebanov-Murugan flow}  \label{limitsec}
 Let us briefly comment about the end point geometries of the  KM flow as this will prove to be useful for the comparison with the flow after NATD.

As indicated above for large radial distances $r\gg a$ the metric in eq. (\ref{eq:KMmetric}) asymptotes the cone metric over $T^{1,1}$, while $H^A_l(r)$ is dominated by its zero mode eq. \eqref{eq:HlPZT}, thus at leading order the warp function becomes
\begin{equation}
L^4 H = \frac{L^4}{r^4} +...\label{GUV}  ,
\end{equation}
which corresponds to the $AdS_5\times T^{1,1}$ UV fixed point solution corresponding to the vacua for which ${\cal U}=0$. In the opposite limit $r\ll a$, deep IR region, the behavior is more subtle. Indeed
close enough to the D3 branes stack, $r\sim 0, \, \theta \sim 0$,  the metric in eq. (\ref{eq:KMmetric}) is given by
\bea
ds^2_{6}=\frac{2}{3}dr^2+d(a\theta)^2+(a\theta)^2d\varphi^2+\frac{r^2}{6}\left(\omega_1^2+\omega_2^2+\omega_3^2\right)
+..., 
\label{line}.
\eea
which upon introducing the coordinate transformation 
\bea
r=2 \sqrt{6}R \cos\alpha, \qquad a\theta =4R\sin\alpha\label{coordtransf}
\eea
the metric in eq. (\ref{line}) becomes, locally, the cone metric over $S^5$; 
$4^2dR^2+4^2R^2ds_{S^5}^2$
Moreover, following the arguments of  \cite{Klebanov:2007us, Martelli:2008cm} the Green's function will be approximated by,
\beq
L^4 H= \frac{L_{IR}^4}{16 R^4}+..., \label{GIR}
\eeq
where  $4 L_{IR}^4=\pi \alpha'^2 N_{D3}$, which reproduces eq \eqref{eq: Ads5s5}. This shows that  a new throat corresponding to $AdS_5\times S^5$  opens up in the IR.

A natural question to address is what is the fate of physical quantities along this flow. In particular, an observable which can be defined along the entire RG flow is the c-function,  roughly 
 measuring  the degrees of freedom that participate in the dynamics at a given energy. At the end points of the RG flow 
 this c-function coincides with the central charge and therefore is conjectured to satisfy $c_{UV}>c_{IR}$. Following \cite{Klebanov:2007ws,Macpherson:2014eza} 
and according to our normalisations the central charges at the fixed points of the KM solution are 
 \beq
c_{{\cal N}=4}=\frac{1}{4} N_{D3}^2,\qquad  c_{{\cal N}=1}=\frac{27}{64} N_{D3}^2,
\eeq
which ratio clearly satisfies $c_{UV}/c_{IR}= \frac{27}{16}>1$ in agreement with the c-theorem. As we will see below this result
will go through even for the end point geometries of the 
non-Abelian T-dualised
KM solution.

\subsection{Dualisation of the Klebanov-Murugan flow} \label{sec:KMdualprobes}

We now apply the technique of non-Abelian T-duality to the geometry 
in eq.~\eqref{eq:KMmetric} along the directions of the $SU(2)$ 
isometry parameterised by $\omega_i$.  
The technique has already been outlined in a preceding section 
and a set of T-duality rules were given.  For concision, below we simply quote the final result of the dualisation.

\subsection{The non-Abelian T-dual IIA solution}  
 The NATD transformation only affects the 3-d part of the metric in eq.~\eqref{eq:KMmetric} that contains the $SU(2)$-isometry parametrised by $\omega_i$. Following the prescription of previous sections this gives,
\begin{align}\label{ds3zz}
\widehat{ds^2}(\mathcal{M}^3)&=\frac{\alpha'\!\!~^2}{\Delta}\bigg[\alpha'\!\!~^2\rho^2 d\rho^2 +L^4 H A_3^2A_1^2 \bigg(\rho^2 \cos^2\chi d\chi^2+\rho d\rho d\chi \sin2\chi+\\
&~~~~~~~~~~ \sin^2\chi\big(d\rho^2+\rho^2(d\xi+\cos\theta d\varphi)^2\big)\bigg)+L^4A_1^4 H\big(\cos\chi d\rho-\rho d\chi \sin\chi\big)^2\bigg]\nn,
\end{align}
where
\beq
\Delta= L^2 \sqrt{H}\big(\alpha'\!\!~^2 \rho^2 A_1^2 \sin^2\chi+A_3^2(\alpha'\!\!~^2\rho^2\cos^2\chi+L^4H A_1^4)\big).
\label{deltazz}\eeq
As in the case of the dual of $AdS_{5}\times S^{5}$ we have expressed 
the Lagrange multipliers  as polar coordinates.  
The remaining seven-dimensional part of the metric remains unchanged.

The NS 2-form generated by the non-Abelian T-duality, is given by,

\begin{align}\label{eq: B2}
\widehat{B}_2&=\frac{L^2\alpha' \sqrt{H}}{\Delta}\bigg[\alpha'\!\!~^2\rho^2 \cos\chi\sin^2\chi(A_3^2-A_1^2)d\xi\wedge d\rho+\alpha'\!\!~^2\rho^3\sin\chi(A_3^2\cos^2\chi+A_1^2\sin^2\chi)d\xi\wedge d\chi\nn\\
&~~~~~~~~~~~~-A_3^2\cos\theta\bigg(L^4A_1^4\rho H\sin\chi d\chi\wedge d\varphi+\cos\chi(\alpha'\!\!~^2\rho^2+L^4 A_1^4 H)d\varphi\wedge d\rho\bigg)\bigg],
\end{align}
and the dilaton is,
\beq
e^{\hat{\Phi}}= \frac{\alpha'\!\!~^{3/2}}{\sqrt{\Delta}}. 
\eeq
The RR-sector is given by
\begin{align}\label{eq:dualRR}
\widehat{F}_2=\frac{ L^4 A_3 A_1^2}{ \alpha'\!\!~^{3/2} \sqrt{\kappa}}\sin\theta\bigg(A_2^2\kappa\frac{\partial H}{\partial r} d\theta\wedge d\varphi+ \frac{\partial H}{\partial \theta} d\varphi\wedge dr\bigg),\;\;\;\;
\widehat{F}_4= \widehat{B}_2\wedge \widehat{F}_2 .
\end{align}
Let us study now the corresponding quantised charges.
\subsection{Quantised Charges and Large Gauge Transformations}

To begin with, it is clear from eq.~\eqref{eq:dualRR} that there is a non zero D6 Page charge,
\beq
Q_{D6}=\frac{1}{2\kappa_{10}^2T_6}\int_{S^2}\widehat{F}_2= \frac{1}{8} N_{D6}(9a^2+r^2)r^3\int_0^{\pi} \sin\theta \frac{\partial H}{\partial r} d\theta= N_{D6},
\eeq
where we fix $L^4= \frac{27}{2}\alpha'~\!^2 N_{D6}$.
One can calculate the quantity $b_0$ defined 
in eq.(\ref{b0zz}). On the cycle $\varphi= 2\pi-\xi$, $d\rho=0$, 
$d\theta=0$ the NS two form is simply
\beq\label{eq:B2cyc}
\widehat{B}_2=\alpha'\rho \sin\chi d\chi\wedge d\xi,\nonumber
\eeq
which gives $b_0=\frac{\rho}{\pi}$. This suggests, following the 
logic below eq.(\ref{b0zz})---and first proposed in \cite{Lozano:2013oma},
\cite{Lozano:2014ata}--
namely  that on crossing the points $\rho= n\pi$, one
 should perform gauge transformations of the form
\beq
\Delta \widehat{B}_2= -\alpha' n \pi \sin\chi d\chi\wedge d\xi,\nonumber
\eeq
to satisfy the requirement that
\beq
0\leq\frac{1}{4\pi^2}\int_{S^2} \widehat{B}_2 <1.\nonumber
\eeq

We may compute the page charge of D4 branes induced after $n$-large gauge transformation and find the result
\beq
\frac{1}{2\kappa_{10}^2 T_4}\int_{S^2\times S^2}
\big(\widehat{F}_4-(\widehat{B}_2+\Delta \widehat{B}_2)
\wedge \widehat{F}_2\big)= 
n \frac{1}{8} N_{D6}(9a^2+r^2)r^3\int_0^{\pi} \sin\theta \frac{\partial H}{\partial r} d\theta.\nonumber
\eeq
In other words
\beq
Q^{Page}_{D4}=n Q^{Page}_{D6}.\nonumber
\eeq
Just like in the case of $AdS_5\times S^5$ studied previously, 
we generate D4 branes.
The D6 branes can be thought as D4's that polarised under 
the influence of the $B_2$-field. They are supersymmetic when wrapping $\rho$ and $(\rho,\chi,\xi)$ respectively and placed at $r=0$.

\subsection{Limits of the NATD Klebanov-Murugan flow}
In this section we shall identify the end point geometries of the KM flow after dualisation paralleling the arguments given in Section \ref{limitsec}.

The unperturbed geometry at the UV fixed point can be identified by taking the limit $r\gg a$ in the dualised KM solution of eqs. (\ref{ds3zz}). After the limit, one can easily see
that, at leading order, this solution will approach the one of the dualised KW geometry. This background is related to the gravitational duals of ${\cal N}=1$ $T_{N}$ or `Sicilian' theories  first introduced in \cite{Benini:2009mz}  and  carefully studied in \cite{Bah:2012dg, Bah:2011je}.

In the IR, as the duality transformation has acted non trivially on the internal geometry, one should 
not expect that locally the space will look like flat space.
 Moreover, as we approach this fixed point the resulting 
geometry will develop a singularity. 
To be more precise, close to the stack of branes and using the coordinate transformation
in eq. (\ref{coordtransf}) one can prove that, at leading order, the background in eq. (\ref{ds3zz}) approximates that of eq. (\ref{ads5xs5natd}). This solution, as explained in Section \ref{sec:NABThol}, 
corresponds to a background of the Maldacena-Gaiotto type. Therefore we can interpret the flow described by the background in eqs. (\ref{ds3zz})-(\ref{eq:dualRR}) as realising
a deformation of a particular Sicilian CFT  by an operator of dimension two. The field theory flows in the IR to a particular Gaiotto-Maldacena CFT.

Finally, let us study the fate of the central charges of the end point geometries of the dualised KM flow. Indeed, quite generally,  it has been shown that the central
charge of the backgrounds obtained via NATD is an invariant up to a constant term \cite{Itsios:2013wd}. 
For the fixed point geometries after NATD using a 
similar logic to the one described in Section \ref{sec:NABThol} we find central charges
\beq
\hat{c}_{{\cal N}={2}}=\frac{1}{12} N_{D6}^2 N_{NS5}^3, \qquad \hat{c}_{{\cal N}=1}=\frac{9}{64}N_{D6}^2 N_{NS5}^3,
\eeq
which ratio $\hat{c}_{UV}/\hat{c}_{IR}=\frac{27}{16}$ is 
in agreement with the c-theorem. As pointed out in \cite{Itsios:2013wd},
the quotient of central charges before and after non-Abelian T-duality
is invariant.

We will now move to calculate field theory observables of our new Type IIA background. More precisely, we will study baryonic condensates and the axionic strings associated with the baryon symmetry breaking. This was studied in great detail in the papers \cite{Martelli:2008cm}.  Nevertheless, note that the calculations of the papers \cite{Martelli:2008cm}
are based on the configuration being $AdS_5\times X_5$ 
and RR five form (or an $AdS_4\times X_7$ with $F_4$ in M-theory). 
In the next subsection, we will compute observables in
the Type IIA background of eqs. (\ref{ds3zz})-(\ref{eq:dualRR}). 
The structure of the geometry and fluxes is very different, but the matching
of our results with those in  \cite{Klebanov:2007us}, 
\cite{Klebanov:2007cx}, \cite{Martelli:2008cm} 
suggests that the nice Mathematics described
by \cite{Martelli:2008cm} may also be at work in our IIA backgrounds.

\subsection{Comments on  the field theory and its observables}
In this section, we briefly  comment on two observables
 in the field theory dual to our 
new background generated by NATD in eqs.(\ref{ds3zz})-(\ref{eq:dualRR}).
The first of them, the baryonic condensate, was originally analyzed in \cite{Gubser:1998fp}. The idea was used in \cite{Klebanov:2007us} 
to study the one point function of  a particular baryonic operator
 in the field theory dual 
to the warped resolved conifold prior 
to non-Abelian T-dualisation ---see also \cite{Martelli:2008cm} 
for a general discussion. 
The second observable, the axionic strings, that appear due to the existence of a Goldstone boson associated with the baryonic
symmetry breaking,  was studied in
\cite{Klebanov:2007cx}.  Below, we find the objects that
represent these QFT observables in our type IIA background.

\subsubsection{Baryonic condensates}
In \cite{Klebanov:2007us} it was shown that the baryonic operator VEV's correspond to D3 branes wrapping
the $ \mathbf{R}^4$ bundle of the resolved conifold. Here we will propose that such baryonic operators correspond
to D0 branes which extend in the radial direction, $r$, 
in the range $[r_0, r_\Lambda]$ (a UV-cutoff, that should be supplemented by
the usual substraction procedure) 
and a D2 brane which, in addition to their radial extent, 
wrap the $S^2$ spanned by $(\chi,\xi)$ .  The D0 branes will be placed at arbitrary
fixed values of $(\theta,\varphi)$ with  $\rho=0$. For the D2 branes, things are little more subtle as large gauge transformations are relevant for this object.

The corresponding induced 1d metric and dilaton for the D0 brane are
\bea
\widehat{ds}_{D0}^2= \frac{L^2 H^{1/2}}{\kappa(r)} dr^2,\;\; \;\;\;\;
e^{-\hat{\Phi}}=\frac{\sqrt{\Delta}}{\alpha'^{3/2}}\Big \lvert_{\rho=0}=
\frac{L^3 H^{3/4}  A_1^2 A_3}{ \alpha'^{3/2}},\;\;\; T_{D0}=\frac{1}{\alpha'^{1/2}}. 
\eea
Evaluating these in the DBI action we get
\bea
 &&S_{D0}= - T_{D0}\int_{r_0}^{r_\Lambda} e^{-\hat\Phi} dr 
\sqrt{\det[\hat{g}_{D0}]},\nonumber\\
&&\;\;\;\;\;\;\;=\left(\frac{3N_{D6}}{4}\right)  \int_{r_0}^{r_\Lambda}dr  r^3 \sum_{l=0}^{\infty}(2l+1)H^A_l(r)P_{l}(\cos\theta) .
\label{actiond0xx}
\eea
We can compare this result with the one obtained in 
equation (6) of \cite{Klebanov:2007us}, 
and conclude that the rest of the calculation for $e^{-S_{D0}}$ will go 
exactly
as in \cite{Klebanov:2007us}. Indeed, the 
dimension of the baryon field after dualisation is
$\Delta=\frac{3N_{D6}}{4} $. 

For the D2 brane, things are more complicated. This is because it extends along $r$ and wraps $(\chi,\xi)$ which means the NS 2-form contributes to its action. We choose to place the D2 at an arbitrary point on $\rho$ within a cell of length $\pi$ {\it i.e.} 
$\rho~\epsilon~ [n\pi,~(n+1)\pi)$, as discussed earlier, this will require a large gauge transformation that will send $\widehat{B}_2\to \widehat{B}_2+\Delta \widehat{B}_2$, which gives
\beq
 \widehat{B}_2+\Delta \widehat{B}_2 = \alpha' \sin\chi\left(\frac{L^2 \rho^3\alpha'~\!^2\sqrt{H}}{\Delta}\left(A_1^2\sin^2\chi+A_3^2\cos^2\chi\right)-n\pi\right)d\chi\wedge d\xi,
\eeq
while the induced metric on the world volume of the D2 is
\beq
ds_{D2}^2=L^2\sqrt{H}\left( \frac{dr^2}{\kappa}+ \frac{\alpha'~\!^2 L^2 \rho^2 A_1\sqrt{H}}{\Delta}\left(A_1^2\sin^2\chi d\chi+A_3^2\left(\cos^2\chi^2+\sin^2\chi d\xi^2\right)\right)\right).
\eeq
One can then compute the DBI action, which turns out to be rather complicated except at $\rho = n\pi$, which is an extremum of the integrand, and leads to 
\beq
e^{-\hat{\Phi}}\sqrt{\det[g_{D2}+\widehat{B}_2+\Delta \widehat{B}_2]}\Big \lvert_{\rho=n\pi} = \frac{L^4 n \pi r^3 H}{18 \sqrt{\alpha'}}\sin\chi \ , 
\eeq
and so the DBI action gives
\bea
&&S_{D2,n}= - T_{D2}\int_{r_0}^{r_\Lambda}\int_{S^2} 
e^{-\hat\Phi} drd\chi d\xi \sqrt{\det[g_{D2}+\widehat{B}_2+\Delta\widehat{B}_2]}\Big \lvert_{\rho=n\pi} \nonumber\\
&&\;\;\;\;\;\;\;=n \left(\frac{3  N_{D6}}{4}\right)  \int_{r_0}^{r_\Lambda}dr  r^3 \sum_{l=0}^{\infty}(2l+1)H^A_l(r)P_{l}(\cos\theta) .
\eea
Thus the D2 branes give rise to a non zero baryonic VEV 
at each point $\rho= n\pi$ for $n>0$ and the dimension 
of the corresponding baryon field is $\Delta_n=n 
\frac{3N_{D6}}{4} = \frac{3N_{D4}}{4}$. 
This gives a total of $(n+1)$ baryonic vevs, one coming 
from the D0 and $n$ coming from D2 branes.

One can also check that the D0 brane is SUSY at $\rho=0$ and also are
the D2 branes precisely at $\rho=n\pi$. To do this one can use 
the $SU(2)$-structure calibration form
\beq
\Psi_{cal}=-8e^{-\hat{\Phi}-A}\text{Im}\hat{\Psi}_{-}\wedge e^{-\hat B_2- \Delta\hat B_2}\nn
\eeq
which may be extracted from Section \ref{sec:Tdualansatz} and Appendix \ref{sec:backgrounds}. The calculation amounts to showing that
\beq
\Psi^{(1)}_{cal}\Big\lvert_{\rho=0} = e^{-\hat{\Phi}}\sqrt{\det[g_{D0}]}\Big\lvert_{\rho=0}dr,~~~~~\Psi^{(3)}_{cal}\Big\lvert_{\rho=n\pi} = e^{-\hat{\Phi}}\sqrt{\det[g_{D2}+\hat B_2+\Delta \hat B_{2}]}\Big\lvert_{\rho=n\pi}dr\wedge d\chi\wedge d \xi,\nn
\eeq
where the superscript refers to the form degree. 
Let us discuss axionic strings following  the treatment in \cite{Klebanov:2007cx}.

\subsubsection{Axionic strings}
In \cite{Klebanov:2007cx}, 
the authors study the presence of 
axionic strings, namely objects that couple to
the Goldstone mode that follows the spontaneous 
baryonic symmetry breaking.
The authors of \cite{Klebanov:2007cx}
proposed that the dynamics of the 
axionic string is given by the DBI action of a D3 brane placed at $r=0$
which wraps the $S^2$ of the manifold
\beq
\Sigma_4=[t,x_1, \theta,\varphi].\nonumber
\eeq
This leads to
\beq
T_{axion,D3}=\frac{a^2}{2 \pi\alpha'\!\!~^2 g_s },\nonumber
\eeq
In Type IIB. We  propose  that  in our Type IIA background, the 
axionic string is represented by
a D4  wrapped on $\Sigma_5=[t,x_1, \theta,\varphi, \rho]$ 
The Lagrangian density of the DBI term of such a brane is 
\beq
\label{eq:LaxionD4}
L_{axion, D4}= e^{-\hat{\Phi}}\sqrt{-\det(g+B)}=\frac{A_2}{\sqrt{\alpha'} }\sqrt{ A^2\Xi_1 \cos^2\theta+A_2^2\Xi_2 \sin^2\theta},
\eeq
where
\beq
\Xi_1= \alpha'\!\!~^2\rho^2+L^4A_1^4 H \cos^2\chi,\qquad \Xi_2=\alpha'\!\!~^2\rho^2+L^4A_1^2 H\big(A^2\sin^2\chi+A_1^2\cos^2\chi\big).\nonumber
\eeq
We fix the $\chi$ dependence by minimising eq.~\eqref{eq:LaxionD4} which leads to $\chi= p \pi/2$ for some integer $p$, we find
\beq
T_{axion,D4}=\frac{a^2N_{NS5}^2}{8 \pi \alpha'\!\!~^2  },
\eeq
where the $N_{NS5}^2$ is due to integrating in $0<\rho<(n+1)\pi$ and we use that $r^4 H(r,\theta)\to 0$ away from $\theta=0$.

In this way we close this brief analysis of two interesting  dual field theoretical observables in our new Type IIA backgrounds. We present now some summary and conclusions for this work.

\section{Summary and Conclusions.}
In this work we elaborated on various aspects of the application of non-Abelian T-duality on string backgrounds with
well-established holographic field theory duals.

To begin with, we discussed new aspects of the case of $AdS_5\times S^5$. We achieved an improved understanding
of the range of the T-dual coordinates and  a sharp expression for the holographic central charge, connecting
our examples with others already studied in the bibliography.

Then, we presented a set of general and  powerful
formulas showing that when non-Abelian T-duality is applied on a large class of 
background (of relevance to holography), it acts as 
a solution generating technique.
This result heavily used the formalism of $SU(2)$ and $SU(3)$-structures existent for
backgrounds with ${\cal N}=1$ SUSY in four dimensions. The existing literature 	\cite{Kelekci:2014ima,Grana:2005sn,Lust:2004ig,Gauntlett:2005ww} does already imply the solution generating nature of the duality when performed on our ansatz, but in addition to explicitly presenting the dual G-structure,  we have provided an alternative proof confirming the result of \cite{Kelekci:2014ima} in this case. 

Finally, as an application of the material developed above, we studied the non-Abelian T-duality of the Klebanov-Murugan background,
generating a type IIA (or M-theory) background describing a very interesting flow.  The end points of this flow are supergravity solutions that are related to the gravity duals of ${\cal N}=1$ and ${\cal N}=2$ $T_N$ CFTs in the UV and IR respectively.  Different aspects of the field theory dual to our
new Type IIA configuration have been discussed. Various technical appendixes complement the presentation.

This work opens the field to many possible future developments. 
At present, the most interesting avenues to pursue are described below.

To begin with, further developing the map between Gaiotto field theories
and backgrounds obtained using non-Abelian T-duality.
In this line, to find, if possible, other Gaiotto-Maldacena backgrounds
using non-Abelian T-duality on a given seed configuration.

It would be interesting to explore in more detail
the holographic understanding of the $\rho$-coordinate and better explain
its range.
For example, if its non-compactness implies that we 
are dealing with a QFT in 4+1 dimensions. Also, it would be good
 to understand
better the Myers effect we are describing in Section \ref{sec:NABThol}
for different intervals  in the 
$\rho$-coordinate.

Our result for the central charge $c\sim N_{D6}^2 N_{NS5}^3$ suggest that
we are dealing with QFT similar to those recently
studied in \cite{Gaiotto:2015usa}
\cite{Gaiotto:2014lca}. It would be important to make this correspondence
sharper.

On the geometry side, it is worth extending the derivation of the T-dual G-structure given in Sections \ref{sec:Ansatz}-\ref{sec:Tdualansatz}
to other cases. For example to include
cascading theories, where the fields $F_3,H_3,\Phi$ are present in the 
seed solution or situations with sources in the seed solution
(corresponding to QFT with flavors \cite{Nunez:2010sf}). Indeed calibrated smeared sources, were not considered in \cite{Kelekci:2014ima}. An extended G-structure derivation would be a necessary step to prove that non-Abelian T-duality is a solution generating technique for supersymmetric $R_{1,3}\times M_6$ in the presence of such sources, using the results of \cite{Koerber:2007hd,Martucci:2005ht}. It is likely that the criteria for SUSY to be preserved is the existence of a set of $N=1$ pure spinors $\Psi_{\pm}$ supported by the seed solution that are independent of the $SU(2)$ coordinates in the frames of eq. \eqref{eq:ein}. A similar condition on the Killing spinor is required for unbroken supersymmetry.

Of slightly different geometrical interest would be to apply our results,
summarised in the table of Appendix \ref{sec:backgrounds}
  to other backgrounds.
Also, the $SU(2)_{diag}$-background
obtained applying non-Abelian Duality to the
KW solution was not studied in the bibliography. It may present very
interesting holographic properties, of interest to physicist working
on ${\cal N}=1$ $T_N$ theories. In this same line, application of our 
formalism to the background proposed by the authors of
\cite{Halmagyi:2004jy} is suitable to provide us with the holographic
version of an RG flow interpolating between ${\cal N}=2$  and ${\cal N }=1$
$T_N$ SCFT as one flows from the UV to the IR. Of similar interest is applying non-Abelian T duality to the background in \cite{Halmagyi:2005pn}, however this would require a generalisation of the ansatz of Section \ref{sec:Ansatz}.

We hope to study these and other topics in forthcoming publications.

\section*{Acknowledgements} 
Various colleagues helped us to improve the contents of this paper with their comments and discussions. We would like to thank: Yago Bea-Besada, Nikolay Bobev, Georgios Itsios, Yolanda Lozano, 
Fidel A. Schaposnik,  Diego Rodriguez-Gomez, Eoin \'O Colg\'ain,  Kostas 
Sfetsos,
Alessandro Tomasiello.

N.T.M is supported by INFN and by the European Research Council under the European Union's Seventh Framework Program (FP/2007-2013) - ERC Grant Agreement n. 307286 (XD-STRING).
C.N is Wolfson Fellow of the Royal Society.  The work of D.C.T was supported in part by FWO-Vlaanderen
through project G020714N and postdoctoral mandate 12D1215N, by the Belgian Federal
Science Policy Office through the Interuniversity Attraction Pole P7/37, and by
the Vrije Universiteit Brussel through the Strategic Research Program``High-Energy
Physics". S. Z. was partially supported by Greek state scholarships foundation (IKY) and PNPC-CONACyT 2014 projects. S. Z.
would also  like to thank the Physics department of the University of Athens for hospitality.

\begin{appendix}

\section{Further comments on the non-Abelian T-dual of $AdS_{5}\times S^{5}$}
\label{appendixpotentials}
\subsection{A comment on the potentials $V_{ST}$ and $V_{app}$}
It may seem a reason to worry that the potential obtained in \cite{Sfetsos:2010uq} is different from the approximate one
obtained in a  series expansion close to $\sigma\sim\eta\sim 0$ on potential in eq.(\ref{vapp}), for the solution in \cite{Maldacena:2000mw}. We will analyse this more closely below.

To establish the comparison, it is convenient to work 
in eleven dimensions. 
We will uplift the solution in eq.(\ref{metrica}). We will use the well-known
formulas,
\bea
& & ds_{11}^2= e^{-2\Phi/3}ds_{10}^2 + e^{4\Phi/3}(dx_{11}+A_1)^2,\nonumber\\
& & C_{3,M}= B_2\wedge (dx_{11}+A_1),\;\; F_{4,M}= dC_{3,M}= 
B_2\wedge F_2 + H_3\wedge (dx_{11}+A_1).\nonumber
\eea
After a short calculation, using that $\Phi_0$ is the constant 
value of the dilaton and
rescaling 
\beq
x_{11}= 2\mu^4 \sqrt{\alpha'}x_{11}\nonumber 
\eeq we find,
\bea 
& & ds_{11}^2= \frac{\alpha' e^{-2\Phi_0/3}}{\mu^2}
(\frac{\dot{V} \Delta}{2 V''})^{1/3}\Big[ 4AdS_{5,\mu} +\mu^2 d\Sigma_5  
+8e^{2\Phi_0}\mu^{14}(\frac{2\dot{V} -\dot{\dot{V} } }{  \dot{V}\Delta   })
(dx_{11}-\frac{2\dot{V}\dot{V'}}{2\dot{V} -\dot{\dot{V} } } d\beta)^2
\Big],\nonumber\\
& & C_{3,M}=2\mu^6 \alpha'^{3/2} 2(\frac{\dot{V}\dot{V'}}{\Delta}-\eta)
d\Omega_2\wedge 
(dx_{11}- \frac{2\dot{V}\dot{V'}}{2\dot{V} -\dot{\dot{V} } } d\beta).
\eea
We now choose conveniently $e^{-2\Phi_0}=4\mu^{12}$ and use that the
eleven dimensional Newton constant is related to the string tension as
$\kappa^{2/3}=(\frac{\pi}{2})^{2/3}L_{P}^2=4^{1/3} \mu^4\alpha'$. 
For the case of the 
IIA 
background in eq.(\ref{ads5xs5natd3}) we can find
a solution to the supergravity 
approximation of M-theory \cite{Gaiotto:2009gz} reads,
\bea 
& & {ds_{11}^2}=\frac{{\kappa^{2/3}}}{\mu^2}
\Big( \frac{\dot{V}\Delta}{2 V''}\Big)^{1/3}  
\Big[  4AdS_{5,\mu} + \frac{2\mu^2 V'' \dot{V}}{\Delta}d\Omega^{2}_2(\chi,\xi)
+ \frac{2\mu^2 V''}{\dot{V}}(d\sigma^2+d\eta^2) +
\frac{4\mu^2 V''}{2\dot{V}-\dot{\dot{V}}}\sigma^2
d{\beta}^2 + \nonumber\\
& & \frac{2\mu^2(2\dot{V}-\dot{\dot{V}})}{\dot{V}\Delta}(dx_{11}+ 
\frac{2\dot{V}\dot{V}'}{2\dot{V}-\dot{\dot{V}}} d{\beta})^2 \Big].\nonumber\\
& & C_3=2\kappa\Big[-\frac{\dot{V}^2 V''}{\Delta} d{\beta}
+(\frac{\dot{V} \dot{V}'}{\Delta}-\eta)dx_{11}
\Big] \wedge d\Omega_2.
\label{mtheorymg}
\eea
We can define an 'interpolating potential'
\beq
V_{int}= \eta\big(\log(\frac{\sigma}{k}) -\frac{\sigma^2}{2k}\big)+\frac{\eta^3}{3k}.
\eeq
Such that for $k=1$, we get $V_{ST}$ and for $k=2$ we get the approximate potential. We will calculate the components of the metric for a 
generic value of $k$. We obtain,
\bea
& & \Big( \frac{\dot{V}\Delta}{2 V''}\Big)=\frac{(k^2-\sigma^2)}{4k^2} (k^2+4k\eta^2-2k \sigma^2+\sigma^4),\;\;\; \frac{2V'' \dot{V}}{\Delta}=\frac{4\eta^2(k-\sigma^2)}{k^2+4k\eta^2-2k \sigma^2+\sigma^4}.\nonumber\\
& & \frac{2V''}{\dot{V}}= \frac{4}{k-\sigma^2},\;\;\; \frac{4V''}{2\dot{V}-\dot{\dot{V}}}\sigma^2=\frac{4\sigma^2}{k},\;\; \frac{2(2\dot{V}-\dot{\dot{V}})}{\dot{V}\Delta}=\frac{4k^3}{(k-\sigma^2)(k^2+4k\eta^2-2k \sigma^2+\sigma^4)}
\nonumber\\
& & \frac{2\dot{V}\dot{V}'}{2\dot{V}-\dot{\dot{V}}}=\frac{(k-\sigma^2)^2}{k^2}, \\
& & \frac{\dot{V}^2 V''}{\Delta} =\frac{8\eta^3(k-\sigma^2)^2}{k(k^2+4k\eta^2-2k \sigma^2+\sigma^4)}, \;\;\;(\frac{\dot{V} \dot{V}'}{\Delta}-\eta)=-\frac{-8 k \eta^3}{k^2+4k\eta^2-2k \sigma^2+\sigma^4}
.\nonumber\label{scalings}
\eea
Given these 'scalings' with the parameter $k$, we observe that if we change variables $\eta\to \sqrt{k}\eta$ and $\sigma\to\sqrt{k}\sigma$, all the terms in the metric are invariant (the overall global factor scales like $k^{1/3}$), while
the last two terms, that enter in the definition of the $C_3$ field scale as $k^{1/2}$. Hence, we could rescale the Newton constant in eleven dimensions
\beq
\kappa\to \kappa \sqrt{k};
\eeq
so that both in the metric and in the $C_3$ these global 'scalings' are absorbed. These shows that both solutions are the same.

\subsection{An interesting operator and cells of the $\rho$-coordinate}
Let us now display two calculations that will add support to our interpretation
of the $\rho$ coordinate presented in Section \ref{sec:NABThol}, 
its range and division in cells of size $\pi$.
Below, we work mostly with the coordinates $(\eta,\sigma)$ related to $(\rho,\alpha)$ by eq.(\ref{cambio}).

We will compute first the mass of the
operator discussed by Gaiotto and Maldacena
around eq.(2.10) of their paper \cite{Gaiotto:2009gz}. 
We start by proposing that in the coordinates 
of eq.(\ref{mtheorymg}), the operator
is represented by a M2 brane that extends in the three-space 
{$\Sigma_3=[t,x_{11}, \eta]_{\sigma=0}$}. 
This is the same set of coordinates used in eq.(\ref{ads5xs5natd3}) for the 
Type IIA background. The three-cycle is placed at $\sigma=0$ and $R=R_0$.

We then calculate the induced metric in M-theory,
\beq
ds^2_{ind}=\frac{\kappa^{2/3}}{\mu^2}
(\frac{\dot{V}\Delta}{2V''})^{1/3}
\Big[ -4R_0^2dt^2 + 
\frac{2\mu^2(2\dot{V} -\dot{\dot{V}})}{\dot{V}\Delta}
dx_{11}^2 +2\frac{V''}{\dot{V}}d\eta^2   \Big].
\eeq
computing the Action of this M2 brane and using that it is equal to the product
of its Energy $E$ and the time interval elapsed $\tau$, we have
\beq
S=T_{M_2}\int_{\Sigma_3}\sqrt{-det g_{ind}}= E \times \tau= \frac{2\pi}{\mu} 
L_{x_{11}}
T_{M2} \kappa R_0 \tau
\int_{0}^{(n+1)\pi/2} d\eta\sim (n+1).
\label{massbps}
\eeq
So, we observe that the mass of the BPS operator is proportional to the range of integration
of the $\eta$-coordinate. This is proportional the range of the $\rho$-coordinate 
according to eq.(\ref{cambio}). As we discussed this integral should be, after crossing
the position $\rho=(n+1)\pi$, the 
same as the number of crossed NS-five branes in the IIA 
picture. This is then the number of
M5 branes in the M-theory picture. Hence the dimension 
or mass of the BPS operator
is proportional to the number of M5-branes, 
$\Delta\sim (n+1)$. This coincides with the result of \cite{Gaiotto:2009gz}
if we interpret the range of the $\eta$ or $\rho$-coordinates as we explained in Section
\ref{sec:NABThol}, rendering support to our proposal.

The second point we want to briefly discuss is the calculation of the number of
M5-branes directly in eleven dimensions. We will integrate $F_4=dC_3$ given by
eq.(\ref{mtheorymg}). The four cycle on which we will integrate is given by
$\Sigma_4=[x_{11},\eta,\chi,\xi]_{\sigma=1}$, that is the cycle
is sitting on the singularity of the background at $\sigma=1$ (or  $\alpha=\frac{\pi}{2}$). 
We calculate the four form
in this submanifold and we obtain,
\beq
F_4|_{\Sigma_4}= 2 \sin\chi d\chi\wedge d\xi\wedge dx_{11}\wedge d\eta.
\eeq
The integral we want to consider is
\beq
T_{M_5}\int_{\Sigma_4}F_4=2T_{M_5}\int_{0}^\pi \sin\chi d\chi \int_{0}^{2\pi} d\xi
\int_{0}^{L_{11}}dx_{11}\int_{0}^{(n+1)\pi/2} d\eta\sim (n+1).
\eeq
We observe then that in coincidence with the point made around eq.(\ref{massbps})
above and in Section \ref{sec:NABThol}, the range of the $\eta$ or $\rho$-coordinates 
should be associated with the number of NS-five branes in the IIA picture or
M5-branes in the eleven dimensional one. As above, 
this computation supports our
proposal of Section \ref{sec:NABThol}. 
It should be interesting to extend this sort
of calculations to geometries that do not admit an electrostatic description
\cite{Petropoulos:2013vya}.
\begin{center}
\begin{landscape}
 \section{Backgrounds falling into the general ansatz} \label{sec:backgrounds} 
  \begin{table}[H]
   \renewcommand{\arraystretch}{1.6}
  \caption{A compendium of $AdS_5$ supergravity backgrounds within the ansatz use in section \ref{sec:Ansatz}  } 
  \vskip 0.2cm
    \begin{tabular}{ | l | c | c | c | c | c | c | c | c | c | }
  \hline
  ~ & $A$ & $\l_1$ & $\l_2$ & $\l_3$ & $h^1$ & $h^2$ & $h^3$ & ${\cal A}_3$ & $ \theta_-$\\  \hline
    \hline
a)~$T^{1,1}$ w.r.t. $SU(2)_{L}$ 	& $\log r$ & $\frac{1}{\sqrt{6}}$ & $\frac{1}{\sqrt{6}}$ & $\frac{1}{3 }$ & $\frac{1}{\sqrt{6}}\sin\theta d\phi$ & $\frac{1}{\sqrt{6}}d\theta $ & $\frac{dr}{r} $ & $\cos\theta d\phi$ & $0$ \\	\hline
b)~ $T^{1,1}$ w.r.t. $SU(2)_{diag}$ & $\log r$ & $\frac{\cos\theta}{\sqrt{3}}$ & $\frac{1}{\sqrt{3}}$ & $\frac{\sqrt{f}}{3\sqrt{2}}$ & $-\frac{\sqrt{8}\sin\theta d\phi}{\sqrt{3} \sqrt{f}} $ & $\frac{\sqrt{\frac{2}{3}}(2r\cos\theta d\theta+3 \sin\theta dr)}{r\sqrt{f}}$ & $\frac{\sqrt{2}(2 \cos\theta dr-r\sin\theta d\theta)}{r\sqrt{f}} $  & $8 \frac{\cos\theta}{f} d\phi$& $2 \phi$ \\	\hline 
c)~KM flow of \cite{Klebanov:2007us} 	& $-\frac{1}{4}\log H$ & $H^{1/4}A_1 $& $H^{1/4}A_1$ & $H^{1/4}A_3$ & $H^{1/4}A_2\sin\theta d\phi$ & $H^{1/4}A_2d\theta $ & $H^{1/4}\frac{dr}{\kappa} $ & $\cos\theta d\phi$& 0 \\\hline
d)~KW flow of \cite{Halmagyi:2004jy}   & $\log H_0$ & $\frac{H_3}{H_0}$ & $\frac{H_4}{H_0}$ & $\frac{v H_5}{H_0}$ & $\frac{u H_1}{H_0}d\phi$ & $-\frac{ H_1}{H_0}du$ & $\frac{ H_2}{H_0}dv $ & $H_6 d\phi$& $\nu \phi$ \\	\hline
e)~$AdS_5\times Y^{p,q}$ &$\log R$& $\frac{1-y}{6}$ & $\frac{1-y}{6}$ & $\sqrt{g} $&$ \frac{1}{3} \sqrt{\frac{q w}{g}} d\a $&$ \frac{(1-y)dy}{3\sqrt{g q w}} +\frac{dR}{6R} \sqrt{\frac{q w}{g}} $& $\frac{dy}{6 \sqrt{g}} - \frac{dR}{3 R}\frac{(1-y)}{\sqrt{g}}$&$ \frac{f w}{g}d \alpha$ &  0 \\ \hline
f)~$AdS_5\times S^5$ &$\log 2R $& $\cos\alpha$ & $\cos\alpha$ & $\cos\alpha  $&$2 \frac{ R \cos\alpha d\alpha +\sin\alpha dR}{R} $&$ 2 \sin\alpha\, d\beta $& $2\frac{ \cos\alpha dR -  R \sin\alpha d\alpha}{R}$& 0 &  $\beta$ \\\hline
  \end{tabular}
  \vskip 0.3cm
Row a. presents the $AdS_5\times T^{1,1}$ background adapted to the $SU(2)_L$ isometry group; Row b. is the $AdS_5\times T^{1,1}$ background adapted to the diagonal $SU(2)_{diag}$ isometry group   in which we define $f(\theta)= 7+ \cos 2\theta$. Row c. is the Klebanov Murugan flow where the functions $A_{1}(r), A_{2}(r),A_{3}(r)$ and $\kappa(r)$ entering into the resolved conifold metric are defined in eq.~\eqref{eq:alpha} and $H(r,\theta)$ obeys a Laplace equation given in eq.~\eqref{eq:diffH}; Row d. gives the ansatz of \cite{Halmagyi:2004jy}   for the Klebanov Witten flow in which $H_{i}(u,v)$ for $i=1\dots 5$ obey a set of BPS equations (see section 4 of \cite{Halmagyi:2004jy})  and can be determined in terms of a single function obeying a quasi-linear, second order PDE ``master equation'', and $H_{0}(u,v)$ is a warp factor that obeys a similar master equation.   Row e. indicates the $AdS_5\times Y^{p,q}$ geometry of  \cite{Gauntlett:2004hh,Martelli:2004wu} where the functions $f(y),q(y),w(y)$ are defined in \cite{Martelli:2004wu} and $g =\frac{q}{6}+ w f^2$ . For completeness, f) gives one option for embedding $AdS_5\times S^5$ into our ansatz, where the $R_{AdS}=R_{S^5}= 2$.  In all cases constant factors such as $L$ are set to unity.
\\[2mm]
For convenience we recall the full IIB ansatz is given in terms of this data by 
\begin{equation}
\begin{aligned}
&ds^{2}= e^{2A} dx^{2}_{1,3} + \sum_{i=1\dots 3} (h^{i})^{2} +   \sum_{j=1\dots 3} (e^{j})^{2}  \ , \quad e^{i} = \l_{i}(\omega_{i} + {\cal A}_{i})  \ , \quad F_{5}=(1+\star_{10}) (\star_{3}dA)\wedge e^{1}\wedge e^{2}\wedge e^{3} \ , \nonumber\\
 & J = h^3 \wedge e^3+ e^1\wedge e^2 +  h^1 \wedge h^2 \ , \quad  \Omega_h = (h^3 + i e^3)\wedge (e^1 + i e^2 ) \wedge (h^1+ i h^2) \ , \quad {\cal A}_1= {\cal A}_2=0 \ .
\end{aligned}
\end{equation}
 \end{table}
\end{landscape}
\end{center}
\newpage

 \section{Rules for non-Abelian T-duality} \label{sec:NABTrules} 

In this appendix we shall present a relatively simple set of Buscher rules that give the non-Abelian T-dual of a background with a $SU(2)$ isometry that fall within the class described in section \ref{sec:Ansatz}. 

  We perform the non-Abelian T-dualisation on these frame fields and choose a gauge fixing in which the Lagrange multipliers, $v^i$, play the r\^ole of T-dual coordinates.  The T-duality acts only on the frame fields $e^{i}$ introduced in  eq.~\eqref{eq:ein} leaving the remaining  seven-dimensional part of the geometry untouched.    A direct application of the Buscher procedure described in detail in \cite{Itsios:2013wd} shows that these frame fields are T-dualised to    
\begin{equation}
\hat{e}^i_\pm = \frac{1}{\Delta} \left( \mu^i_\pm + \nu^i_\pm   \right) \ , 
\end{equation} 
where 
\begin{equation}
\begin{aligned}
  \Delta &= \det M  =  \l_1^2 \l_2^2 \l_3^2  +  \l_i^2 v_i^2 \ ,  \\ 
 \mu^i_\pm &= \epsilon_{(i)jk} \l_{(i)} \l_j^2 v_j dv_k \mp   v_{(i)}\l_{(i)} v_j dv_j \mp   \l_1^2 \l_2^2 \l_3^2 \frac{dv_{(i)}}{\l_{(i)}} \ ,  \\
  \nu^i_\pm  &= \l_j^2 v_j^2 \l_{(i)} {\cal A}_{(i)} -  \l_j^2 v_j {\cal A}_j \l_{(i)} v_{(i)}  \mp \e_{(i)jk}  \l_1^2 \l_2^2 \l_3^2  \frac{{\cal A}_j v_k}{\l_{(i)}} \ , 
\end{aligned}
\end{equation} 
in which indices in brackets are not summed.  The {\em plus} and {\em minus} frame fields are those seen by left and right movers respectively after duality.  Though they seem formidable, these frame fields   obey nice relations
\begin{align}\label{eq:eidents}
&\sum_{i=1}^3 \l_i v_i \hat{e}^i_\pm = \mp  v_i dv_i,\\[2mm]
& \hat{e}^1_\pm \wedge  \hat{e}^2_\pm \wedge  \hat{e}^3_\pm = \mp \frac{\l_1 \l_2 \l_3 }{\Delta} \left( dv_1\wedge dv_2 \wedge d v_3 + {\cal A}_{[i} v_{j]} \wedge dv_i \wedge dv_j+ \epsilon_{ijk}(v_i v_k dv_i+\frac{1}{2} v_k^2 dv_k)\wedge {\cal A}_i\wedge {\cal A}_j \right) \nn . 
\end{align} 

The T-dual metric on the three-dimensional space in which the duality acts  is given by 
\begin{equation}
\begin{aligned}
\widehat{ds}^2 & = \sum_{i=1\dots 3} \hat{e}_+^i   \hat{e}_+^i  =  \sum_{i=1\dots 3} \hat{e}_-^i   \hat{e}_-^i   = {\cal G}_{ij} dv^i dv^j  +2 {\mathbb  A}_i dv^i  +{\mathbb  A}_i {\cal G}^{ij} {\mathbb  A}_j   \ ,  \\
{\cal G}_{ij}  & = \frac{1}{\Delta}\left( v_i v_j + \frac{\l_1^2 \l_2^2 \l_3^2}{\l_{(i)}^2} \delta_{(i)j} \right)  \ , \quad
{\mathbb A}_i = \frac{\l_1^2 \l_2^2 \l_3^2 }{\Delta}\epsilon_{(i)jk} \frac{{\cal A}_j v_k}{\l_{(i)}^2 } \  .
\end{aligned}
\end{equation}

The T-dual NS two form is given by,
\begin{equation}
\begin{aligned}
\hat{B}&=  \frac{1}{2} \e_{ijk} v_i \frac{\hat{e}_\pm^j\wedge \hat{e}_\pm^k }{ \l_j \l_k} +  \e_{ijk}  v_j \l_k^{-1} {\cal A}_i\wedge\hat{e}_\pm^k +\frac{1}{2}\e_{ijk} v_i {\cal A}_j\wedge {\cal A}_k  \mp \l_i \hat{e}^i_\pm  \wedge {\cal A}_i \ , \\
&= \frac{1}{2\Delta}\left( \epsilon_{ijk} \l_i^2 v_i dv_j\wedge dv_k -\l_1^2\l_2^2\l_3^2\epsilon_{ijk}v_i {\cal A}_j\wedge {\cal A}_k+ 2\l_1^2 \l_2^2 \l_3^2  dv_i \wedge {\cal A}_i +2 \l_j^2 v_j v_i dv_i\wedge {\cal A}_j   \right)
\end{aligned}
\end{equation}
and, as usual, the Dilaton acquires a one-loop shift 
\begin{equation}
e^{-2 \widehat{\Phi} } = e^{-2 \Phi} \Delta \ . 
\end{equation}

Since $\hat{e}_+^i$ and $\hat{e}^i_-$ define the same metric they are related by a Lorentz rotation, 
\begin{equation}
\hat{e}_+ = \Lambda \hat{e}_- \ , \quad \Lambda^{i}{}_{j} = \frac{1}{\Delta} \left( \left(\Delta - 2 \l_1^2 \l_2^2 \l_3^2 \right)\delta_{ij} -2 \l_i v_i \l_j v_j - 2 \l_1 \l_2 \l_3 \e_{ijk} v_k \l_k  \right) \ . 
\end{equation}
This Lorentz rotation gives rise to an action on spinors 
\begin{equation}\label{eq: OmegaMatrix}
\Omega \Gamma^i \Omega^{-1} = \Lambda^i{}_j \Gamma^i \ , \quad \Omega = \frac{\Gamma_{11}}{\Delta} \left(  \l_1 \l_2 \l_3 \Gamma^{123} + \l_i v_i \Gamma^i \right) \ . 
\end{equation}
Using this rotation one derives that the T-dual RR forms
\begin{equation}\label{eq:dualflux}
\widehat F_2 =  \l_1 \l_2 \l_3 {\cal F}_2 \ , \quad \widehat F_4 =   \frac{1}{2}\e_{ijk}  \l_i v_i  {\cal F}_2  \wedge \hat{e}_+^j \wedge \hat{e}_+^k = \widehat{B}_2 \wedge \widehat F_2 +   (dv_i\wedge {\cal A}_i+\frac{1}{2}\epsilon_{ijk}v_i {\cal A}_j\wedge {\cal A}_k )\wedge \widehat{F}_2 \ , 
\end{equation}
together with their Hodge duals, 
\begin{equation}\label{eq:stardualflux}
\begin{aligned}
\widehat F_6&= - \star \widehat{F}_4 =  e^{4A} \star_3 {\cal F}_2  \wedge d^4 x \wedge v_i dv_i     \ ,  \\
\widehat F_8 &= \star \widehat{F}_2 = e^{4A} \l_1 \l_2 \l_3  \star_3 {\cal F}_2    \wedge d^4 x \wedge \hat{e}_+^1 \wedge \hat{e}_+^2 \wedge \hat{e}_+^3= B\wedge \widehat{F}_6-\frac{1}{3!v_i}\epsilon_{ijk}dv_j\wedge dv_k \wedge \widehat{ F}_6 \ .
\end{aligned}
\end{equation}

The frame fields that enter in eq.~\eqref{eq:eidents} are the ones that arise directly by following the Buscher procedure as detailed in the appendix of \cite{Itsios:2013wd} however these are not the simplest ones with which to describe geometries with supersymmetry. Simpler results are obtained   after a supplementary Lorentz rotation  of the frame fields in the basis $\hat{E}=\{  h^{3}, \hat e^{1}_{+},\hat e^{2}_{+},\hat e^{3}_{+}\}$   given by   \cite{Barranco:2013fza}:
  \begin{equation}
{\cal R} = \frac{-1}{\sqrt{1 + |\z|^2 }} \left(   \begin{array}{cccc}
    1 & \zeta^{1}  &  \zeta^{2} & \zeta^{3}\\ 
 - \zeta^{1}& 1&- \zeta^{3} & \zeta^{2} \\ 
 -  \zeta^{2}& \zeta^{3} & 1& -\zeta^{1}\\ 
 -  \zeta^{3}& -\zeta^{2}& \zeta^{1} & 1 \\ 
  \end{array}\right)  \ , \quad \zeta^i = \frac{\l_iv_i}{ \l_1\l_2\l_3} \ . 
\end{equation}
Setting  ${\cal A}_{1}={\cal A}_{2}=0$ (as required by supersymmetry) and computing $\tilde{E}= {\cal R}\cdot \hat{E}$   produces the rather simple result of the frame fields in \eqref{eq:dualframesneat}.

\section{Verification that the Dual $SU(2)$-Structure obeys the Supersymmetry Conditions}\label{appendix:proof}
In this appendix we show explicitly that the $SU(2)$-Structure of Section \eqref{sec:Tdualansatz} obeys the required supersymmetry conditions.

Since the NS 3-form $H_3$ is generically rather complicated we find it easier to work with a modified form of the supersymmetry conditions, namely
\begin{align}
&d(e^{2A-\widehat{\Phi}}\widehat{\Psi}_+\wedge e^{-\widehat{B}})=0,\label{eq:diffmod1}\\[2mm]
&d(e^{A-\widehat{\Phi}}Re\widehat{\Psi}_-\wedge e^{-\widehat{B}})=0\label{eq:diffmod2},\\[2mm]
&d(e^{3A-\widehat{\Phi}}Im\widehat{\Psi}_-\wedge e^{-\widehat{B}})=\frac{e^{4A}}{8}\big(\star_6 \widehat{F}_2-\star_6 \widehat{F}_4\big)\wedge e^{-B}\label{eq:diffmod3},
\end{align}
where $H$ does not appear explicitly\footnote{These are simply $e^{-B}\wedge$ the standard differential supersymmetry condition the pure spinors must satisfy, which may be found in section 4.1 of \cite{Grana:2006kf}.}.
Eq. (\ref{eq:diffmod1})  yields two conditions
\be\label{eq:Psipdualeqs}
d(e^{3 A - \widehat{\Phi}+ i \hat{\theta}_+} \omega) =  0 \ , \quad d(e^{3 A - \widehat{\Phi}+ i \hat{\theta}_+} \omega \wedge (2\widehat B- z\wedge \bar{z} )) = 0   \ . 
\ee
Using  eq.~\eqref{eq:dthetaphi} and eq.~\eqref{eq:como} it is possible to show that
\be
i e^{3A-\widehat{\Phi}+i\widehat{\theta}_+}\omega= d\bigg(e^{3A+i\widehat{\theta}_+}\l_3(\l_1v_2- i\l_2 v_1) (h^1+ih^2)\bigg) , 
\ee
so the first of these is solved trivially. In a similar vein, one can re-express
\begin{align}
i e^{3A-\widehat{\Phi}+i\widehat{\theta}_+} \wedge \omega\wedge (2\widehat B- z\wedge \bar{z} )) =& -2\big({\cal A}_3- i \l_3^{-1} h^3\big)\wedge\bigg[ -\l_3e^{3A+i\widehat{\theta}_+}\big(h^1+i h^2\big)\wedge d\big((\l_1v_2- i\l_2 v_1)\big)+\nn\\[1mm]
 &~~~~~~\big(\l_1v_2- i\l_2 v_1\big)d\big(\l_3e^{3A+i\widehat{\theta}_+} (h^1+ih^2)\big)\bigg]\wedge dv_3.
\end{align}
The term in square brackets is exact so one need only consider $d ({\cal A}_3 - \l_3^{-1} h^3)$ to take the exterior derivative. Clearly this gives zero when wedged into the second square bracketed term, as $\mathcal{M}_3$ is three-dimensional, the remaining term then vanishes once we apply eq.\eqref{eq: weird}.

Moving on the eq. \eqref{eq:diffmod2} we find three further conditions,
 \beq
d\big(e^{2A-\widehat{\Phi}}w\big)=0,~~~
d\big(e^{2A-\widehat{\Phi}}(v\wedge j+ w\wedge\widehat{B})\big)=0,~~~
d\big(e^{2A-\widehat{\Phi}}(w\wedge j\wedge j-w\wedge \widehat{B} \wedge \widehat{B} - 2v\wedge j\wedge \widehat{B})\big)=0.
\eeq
Using eq.~\eqref{eq:ersol} one finds that
\begin{align}
&e^{2 A - \hat{\Phi}} w =- d\big(e^{2A} \l_1\l_2 v_3\big),\\[2mm]
&e^{2A-\widehat{\Phi}}(v\wedge j+ w\wedge \widehat{B})=e^{2A}(h^1\wedge h^2-\l_3 {\cal A}_3\wedge h^3)\wedge v_idv_i-d(e^{2A} \l_1 \l_2 )\wedge dv_1\wedge dv_2-e^{2A}\l_1\l_2\l_3 Vol(\mathcal{M}_3),\nn\\[2mm]
&e^{2A-\widehat{\Phi}}(w\wedge j\wedge j-w\wedge \widehat{B} \wedge \widehat{B} - 2v\wedge j\wedge \widehat{B} )= -2 e^{2A}(h^1\wedge h^2-\l_3 {\cal A}_3\wedge h^3)\wedge d^3v\nn,
\end{align}
where $d^3v= dv_1\wedge dv_2 \wedge dv_3$. These are all closed due to eq.~\eqref{eq:dthphi2} and so eq. \eqref{eq:diffmod2} is satisfied.

All that remains is to show that eq. \eqref{eq:diffmod3} is also solved. It gives rise to the following conditions,
\begin{align}
&d\big(e^{4A-\widehat{\Phi}}v\big)+e^{4A}\star_6 \widehat{F}_4=0\\[2mm]
&d\big(e^{4A-\widehat{\Phi}}(w\wedge j-v\wedge \widehat{B} )\big)-e^{4A}\big(\star_6 \widehat{F}_2+\widehat{B} \wedge\star_6 \widehat{F}_4\big)=0,\nn\\[2mm]
&d\big(e^{4A-\widehat{\Phi}}(v\wedge j\wedge j-v\wedge \widehat{B} \wedge \widehat{B} +2 w\wedge j\wedge \widehat{B})\big)+e^{4A}\big(2\widehat{B}\wedge \star_6 \widehat{F}_2+\widehat{B}\wedge \widehat{B} \wedge\star_6 \widehat{F}_4\big)=0\nn.
\end{align}
First we need to find the flux terms which may be extracted from eq. \eqref{eq:simplefluxes}. We have
\beq
e^{4A}\star_6 \widehat{F}_4 = -d\big(e^{4A} v_i dv_i\big),~~~e^{4A}\big(\star_6\widehat{F}_2+\widehat{B} \wedge \star_6 \widehat{F}_4\big) = -d\big(e^{4A} d^3v\big),~~~\widehat{B} \wedge \widehat{B} \wedge \star_6 \widehat{F}_4= \widehat{B} \wedge \star_6\widehat{F}_2=0,
\eeq
where we use that $\star_3 {\cal F}_2= 4 d A$.
The terms inside  derivatives can be manipulated with eq.~\eqref{eq:ersol} to give
\begin{align}
&e^{4A-\widehat{\Phi}}v = e^{4A} v_i dv_i-\frac{1}{2} d\big(e^{4A}\l_1^2\l_2^2\big),\nn\\[2mm]
&e^{4A-\widehat{\Phi}}(w\wedge j-v\wedge \widehat{B} ) =- e^{4A} d^3 v- e^{2A}d\big(e^{2A} \l_1 \l_2 v_3\big)\wedge \big(h^1\wedge h^2- \l_3 {\cal A}_3\wedge h^3\big),\nn\\[2mm]
&e^{4A-\widehat{\Phi}}(v\wedge j\wedge j-v\wedge \widehat{B}\wedge \widehat{B}+2 w\wedge j\wedge \widehat{B})=-2 e^{4A}\l_3 dv_1\wedge dv_2\wedge Vol(\mathcal{M}_3).
\end{align}
Acting on these with the exterior derivative and using eq.~\eqref{eq:dthphi2} then gives precisely minus the contribution coming from the fluxes, solving eq. \eqref{eq:diffmod3}. This completes the demonstration that the pure spinor conditions are solved after dualisation.  
 
\end{appendix}

\end{document}